\documentclass[preprint,12pt]{elsarticle}

\usepackage{graphicx}
\graphicspath{{figures/}}
\usepackage{float}
\usepackage{amssymb}

\usepackage{xcolor}
\usepackage[utf8]{inputenc}

\usepackage[colorlinks,
    linkcolor={black},
    citecolor={black},
    urlcolor={black}]{hyperref}

\journal{ Computational Materials Science }

\begin{document}
\begin{frontmatter}


\title{AiiDAlab -- an ecosystem for developing, executing, and sharing scientific workflows}

\author[marvel,theos,lsmo]{Aliaksandr V. Yakutovich \textsuperscript{*,}\corref{contrib}}
\author[marvel,empa]{Kristjan Eimre\corref{contrib}}
\author[marvel,empa]{Ole Schütt\corref{contrib}}
\author[marvel,theos,lsmo]{Leopold Talirz}
\author[theos]{Carl S. Adorf}
\author[theos]{Casper W. Andersen}
\author[marvel,empa]{Edward Ditler}
\author[marvel,theos]{Dou Du}
\author[marvel,empa]{Daniele Passerone}
\author[marvel,lsmo]{Berend Smit}
\author[marvel,theos]{Nicola Marzari}
\author[marvel,theos]{Giovanni Pizzi\corref{corauthor}}
\author[marvel,empa]{Carlo A. Pignedoli\corref{corauthor}}

\address[marvel]{National Centre for Computational Design and Discovery
of Novel Materials (MARVEL), \'Ecole Polytechnique F\'ed\'erale de Lausanne, 
CH-1015 Lausanne, Switzerland}
\address[theos]{Theory and Simulation of Materials (THEOS), 
    Facult\'e des Sciences et Techniques de l'Ing\'enieur, 
    \'Ecole Polytechnique F\'ed\'erale de Lausanne,
    CH-1015 Lausanne, Switzerland}
\address[lsmo]{Laboratory of Molecular Simulation (LSMO),
    Institut des Sciences et Ingenierie Chimiques,
    Valais, \'Ecole Polytechnique F\'ed\'erale de Lausanne,
    CH-1951 Sion, Switzerland}
\address[empa]{nanotech@surfaces laboratory,
    Swiss Federal Laboratories for Materials Science and Technology (Empa),
    CH-8600 D\"ubendorf, Switzerland}

\cortext[corauthor]{Corresponding author}
\cortext[contrib]{Authors contributed equally}

\begin{abstract}

Cloud platforms allow users to execute tasks directly from their web browser and are a key enabling technology not only for commerce but also for computational science.
Research software is often developed by scientists with limited experience in (and time for) user interface design, which can make research software difficult to install and use for novices. When combined with the increasing complexity of scientific workflows (involving many steps and software packages), setting up a computational research environment becomes a major entry barrier.
AiiDAlab is a web platform that enables computational scientists to package scientific workflows and computational environments and share them with their collaborators and peers.
By leveraging the AiiDA workflow manager and its plugin ecosystem, developers get access to a growing range of simulation codes through a python API, coupled with automatic provenance tracking of simulations for full reproducibility.
Computational workflows can be bundled together with user-friendly graphical interfaces and made available through the AiiDAlab app store.
Being fully compatible with open-science principles, AiiDAlab provides a complete infrastructure for automated workflows and provenance tracking, where incorporating new capabilities becomes intuitive, requiring only Python knowledge.

\end{abstract}


\end{frontmatter}


\section{Introduction}
\label{intro}
Computational simulations constitute an integral part of contemporary science, and are often referred to as the ``third pillar of science'', after theory and experiments \cite{skuse_third_2019}.
So far, performing simulations has been considered a task for computational scientists, despite being more and more apparent that simulations, as a fundamental bridge between theory and experiment, can benefit a much larger community.
To date, running advanced scientific software requires expert knowledge not only on the phenomena to be modelled, but also on how to access and operate compute resources, develop, compile, and install the relevant software, and how to  optimally use high-performance computing (HPC) resources: from the choice of parallelisation strategy to the most efficient number of computational nodes.
To mitigate these challenges and to offer a common ground for researches, we introduce AiiDAlab~\cite{noauthor_aiidalab_nodate-3,noauthor_aiidalab_nodate-1}, an open-source platform that provides the means to develop, execute, and share computational workflows as intuitive web services.
Exploiting modern web technologies, AiiDAlab grants the capability of working with advanced simulation tools to any interested researcher. 

AiiDAlab is built on top of AiiDA~\cite{pizzi_aiida:_2016,huber_aiida_2020,noauthor_aiida_nodate}, an open-source Python infrastructure developed to help researchers automating, managing, persisting, sharing and reproducing workflows and associated data.
AiiDA has enabled the community to perform high-throughput simulations~\cite{mounet_two-dimensional_2018,noauthor_aiida_nodate-1} and publish the full provenance information of all steps performed~\cite{mounet_two-dimensional_2019}.
In AiiDAlab, AiiDA workflows can be packaged together with user-friendly graphical interfaces for execution and data analysis into accessible applications, and  the platform provides an intuitive environment to manage, run and, publish these applications.
AiiDAlab applications are primarily implemented in Python, leveraging Jupyter~\cite{noauthor_jupyter_nodate} and Ipython widgets~\cite{noauthor_jupyter-widgets/ipywidgets_2019} for the user interfaces.
This enables every computational scientist with Python knowledge to develop and publish applications on the AiiDAlab platform.

The challenges in tackling scientific workflows, data sharing and reproducibility have been addressed by a number of initiatives launched in the last decade. Some of these resulted in the creation of repositories for storage, management, and sharing of scientific data~\cite{villars_pauling_2004,zarkevich_nikolai_structural_2006,jain_high-throughput_2011,curtarolo_aflowlib.org:_2012,saal_materials_2013,landis_computational_2012,noauthor_eudat_nodate,noauthor_nomad_nodate}.
Others focused on developing general formats to store the results computed with different quantum chemistry packages~\cite{adams_quixote_2011,noauthor_open_nodate}.
Finally, a consortium of scientists, industry, funding agencies, and scholarly publishers has formulated the ``FAIR'' principles~\cite{wilkinson_fair_2016} that are supposed to guide the data holders in making their data reusable.

Platforms providing a web-environment to run scientific software already exist. Notably within industry~\cite{koschmieder_aixvipmapoperational_2019,noauthor_medea_nodate,noauthor_citrine_nodate,noauthor_exabyteio_nodate} (and thus closed source) or open access~\cite{noauthor_nanohub.org_nodate}. From 2018 Google made open access the Jupyter based collaborative  platform named Google Colab~\cite{noauthor_google_nodate}.

Here we highlight how AiiDAlab, when coupled with the Materials Cloud~\cite{talirz_materials_2020,noauthor_materials_nodate} dissemination platform, can form an integrated solution for seamlessly creating FAIR~\cite{wilkinson_fair_2016} data. 
Automation allows greatly accelerating human response as a bottleneck in materials design, which occurs whenever the time needed to correctly setup calculations is comparable to the time it takes to execute them.
AiiDAlab allows to easily enable high levels of standardisation in the way calculations are performed and in the way results are analyzed and processed.
In a research laboratory where experimentalists collaborate with computational scientists, the platform allows intuitive and direct access to the submission of calculations and to the analysis of data that are processed automatically.

This manuscript is divided into two main sections: a detailed description of the AiiDAlab architecture and the presentation of three different AiiDAlab applications.


\section{The AiiDAlab platform}
\label{section_architecture}

\begin{figure}[H]
\centering\includegraphics[width=1.0\linewidth]{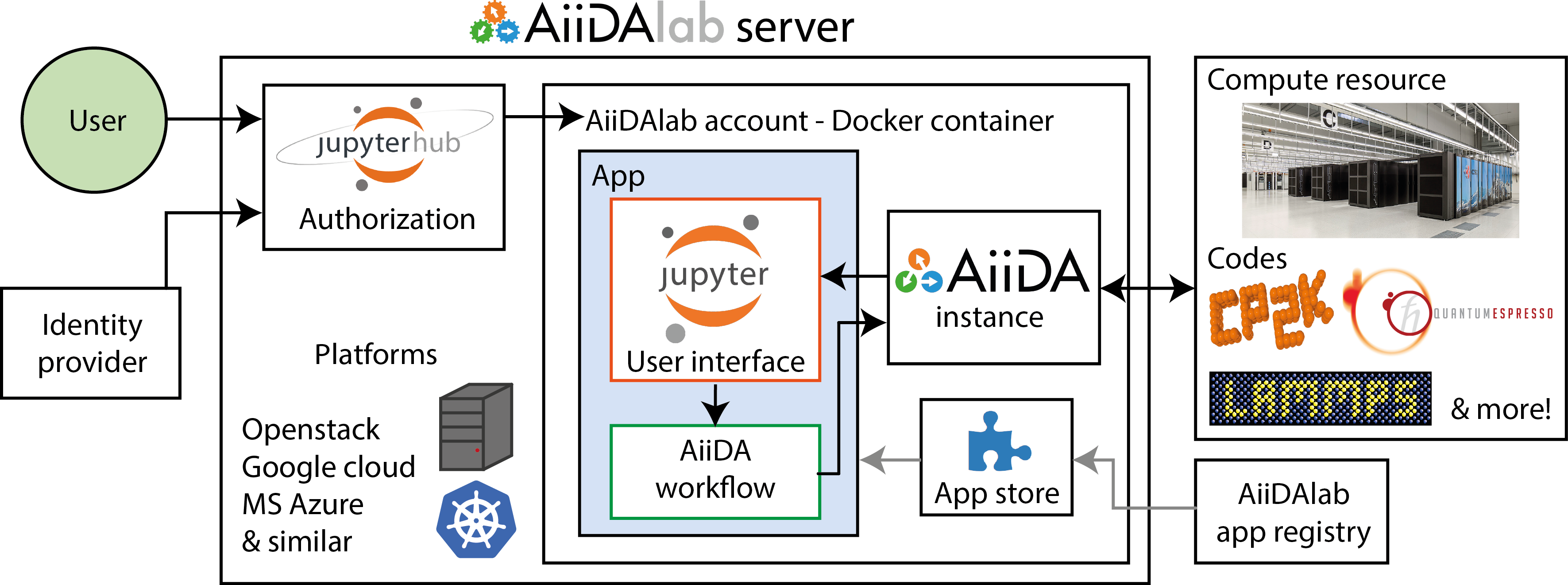}
\caption{Diagram of the AiiDAlab architecture. Users are authorized through JupyterHub (which performs the authentication through an external identity provider service) and assigned to a specific personalized AiiDAlab instance -- a Docker container with an AiiDA instance and a Jupyter-based environment to manage AiiDAlab apps. The apps contain Jupyter-based user interfaces and AiiDA workflows, which communicate with the AiiDA instance. New apps can be installed through the App store app, which is a user interface to the AiiDAlab application registry. App developers can modify the apps through their AiiDAlab account or directly push their changes to the code repository. The AiiDAlab service can be hosted on a Kubernetes cluster, on a conventional cloud server (OpenStack, Amazon Web Services, Google Compute Cloud, Microsoft Azure, etc), or on a local server. Images and logos are reproduced in compliance to the corresponding licenses\cite{noauthor_cscs_nodate,noauthor_cp2k_nodate,noauthor_quantum_nodate-1,noauthor_lammps_nodate-1,noauthor_jupyter_nodate-1,noauthor_jupyterhub_nodate,noauthor_kubernetes_nodate}.}
\label{fig:aiidalab_map}
\end{figure}

The main goal of the AiiDAlab platform is to provide an environment for users with diverse expertise to access and run computational workflows embedded in AiiDAlab apps. The general structure of the platform is shown in Fig.~\ref{fig:aiidalab_map}. Each user has a separate AiiDAlab account, which grants access to their AiiDAlab instance through a web browser. The AiiDAlab instances are Docker containers running on a remote cloud server with a Jupyter-based environment to manage AiiDAlab apps and an AiiDA instance that takes care of communicating with remote computers, executing automatically all steps defined in the application workflows, and retrieving and parsing the output data. The concepts of the platform are described in detail in the following sections.

\subsection{AiiDAlab application}
\label{sub_section_app}
Central to AiiDAlab is the concept of an ``application'' (app): a set of AiiDA workflows that encode simulation tasks, combined with graphical user interfaces (GUIs) to manage them.
The apps allow users to submit calculations to different remote computational facilities, monitor the status of the calculations, analyze the data that are automatically retrieved, and run post-processing calculations.

\begin{figure}[H]
\centering\includegraphics[width=0.9\linewidth]{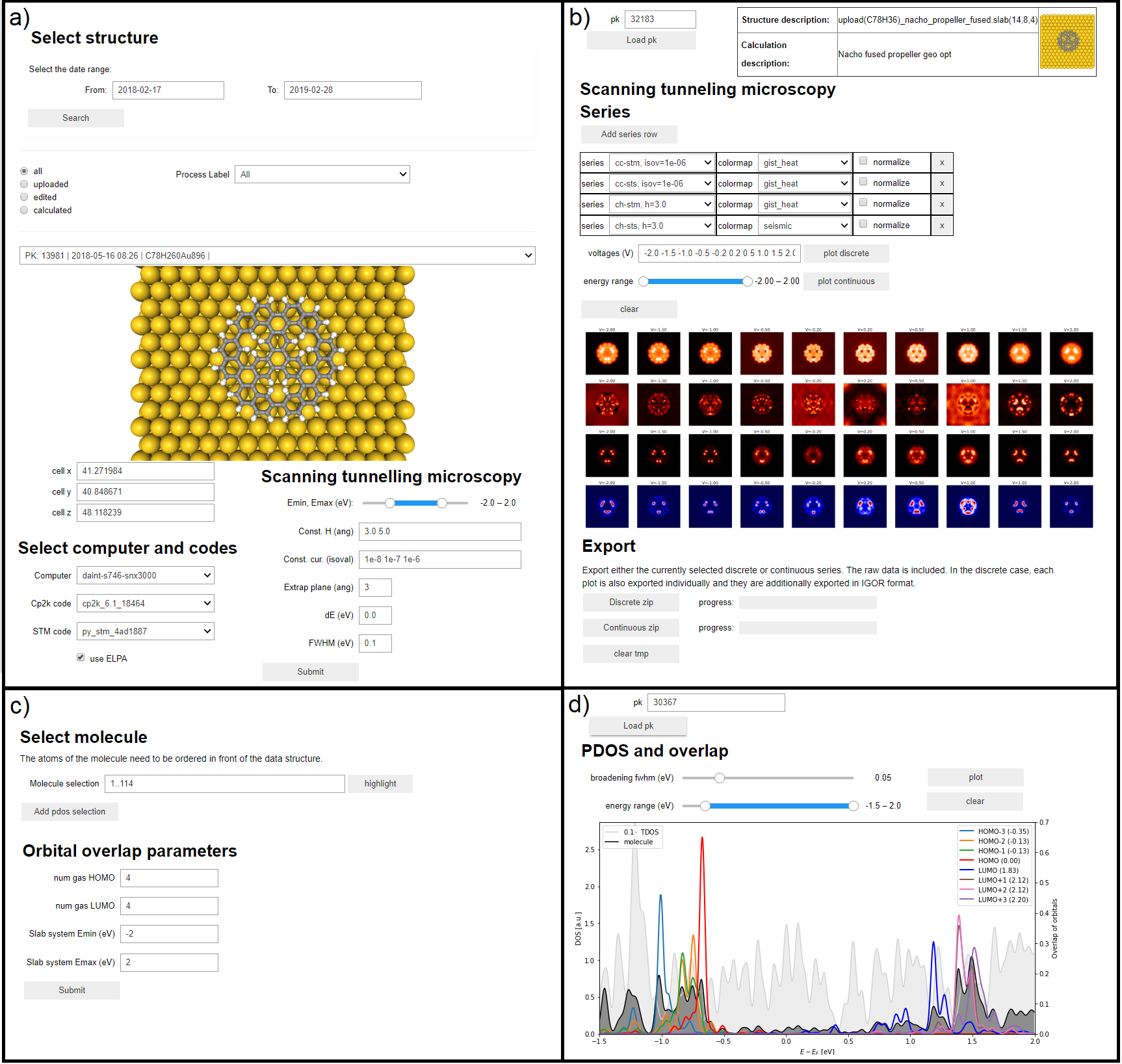}
\caption{Snapshots of the Scanning Probe Microscopy (SPM) app. (a) Interface for the submission of a workflow to compute Scanning Tunnelling Microscopy (STM) images. (b) Interface to analyze the results of the STM calculations. (c) Interface to submit the workflow to compute the Projected Density of States (PDOS) for a system composed by a molecule adsorbed on a substrate. (d) Analysis of the PDOS results. See Fig.~S$10$ for more details.}
\label{fig:spm}
\end{figure}

The core of a scientific AiiDAlab application is constituted by the AiiDA workflows, which encode the computational and scientific knowledge about a simulation.
The development of a reliable computational workflow is a complex design problem.
The scientists developing the workflow are responsible for the methodologies and related approximations that are selected to tackle the scientific problem under investigation.
Additionally, an efficient workflow should be robust and resistant to externalities, such as temporary failures of the remote computing facility, network failures or shutdowns.
Several such cases can be resolved by AiiDA itself; for instance, if a computer becomes unavailable, AiiDA detects it and puts the calculation on hold.
However, some situations are specific to a given simulation code (run time error due to lack of convergence of an interative routine, or a problem with the input parameters).
These cases must be handled by the corresponding workflows.

All user interactions within an AiiDAlab app are facilitated by Jupyter notebooks~\cite{noauthor_jupyter_nodate}, which are used to submit and manage the AiiDA workflows, and then visualize, analyze and post-process their results.
The interactive GUI elements, or ``widgets'', are mainly built using Python and the ipywidgets module \cite{noauthor_jupyter-widgets/ipywidgets_2019}.
Development and maintenance of the Jupyter notebooks is straightforward and can be done directly in the Python code.
To turn notebooks into user-friendly web applications, a Jupyter extension that we developed, named ``Appmode''~\cite{schutt_appmode:_2019} is enabled by default.
This extension hides away all the code from the Jupyter notebooks, leaving only the GUI elements visible to the user.

As an example, Fig.~\ref{fig:spm} shows several screenshots of the graphical user interface of the recently released Scanning Probe Microscopy application, which is designed for the characterization of molecules adsorbed on substrates by means of density functional theory calculations.
The app contains two AiiDA workflows: one to simulate scanning tunnelling microscopy (STM) experiments, and one to investigate how the orbitals of a molecule hybridize with the orbitals of the substrate.
The submission interfaces (Fig.~\ref{fig:spm}a,c) guide the user in specifying a minimal set of input parameters needed to prepare and submit the workflows.
Furthermore, the app includes interfaces to visualize and export the results that have been automatically post-processed, shown in Fig.~\ref{fig:spm}b,d.
This Scanning Probe Microscopy app is discussed in more detail in Section \ref{spm_app}.

For more technical details about AiiDAlab applications, the reader is referred to the ``AiiDAlab application`` section of the Supplementary Information, with information about the structure of the applications, their creation, and examples of usage of Appmode.

Importantly, a library of user-interface elements (widgets) and their logic to interface with the underlying AiiDA infrastructure is available as a base app called ``AiiDAlab widgets''. This app provides out-of-the-box some of the high-level functionalities that are most often needed in AiiDAlab applications.
Examples of such high-level functionalities include uploading data from a local computer, browsing data that are stored in an AiiDA database, and selecting the computational resources where to run the simulations (see more in the ``Monitoring submitted calculations'' section of the Supplementary Information).
This base app is contained in the GitHub repository aiidalab-widgets-base~\cite{noauthor_reusable_2018}; every AiiDAlab user can easily import this module and start using the base widgets capabilities.
An extended documentation containing the available widgets and examples of usage are available in a dedicated web-page~\cite{noauthor_aiidalab_nodate-2}.

\subsection{AiiDAlab home app}
\label{subsection_home}
\begin{figure}[H]
\centering\includegraphics[width=1\linewidth]{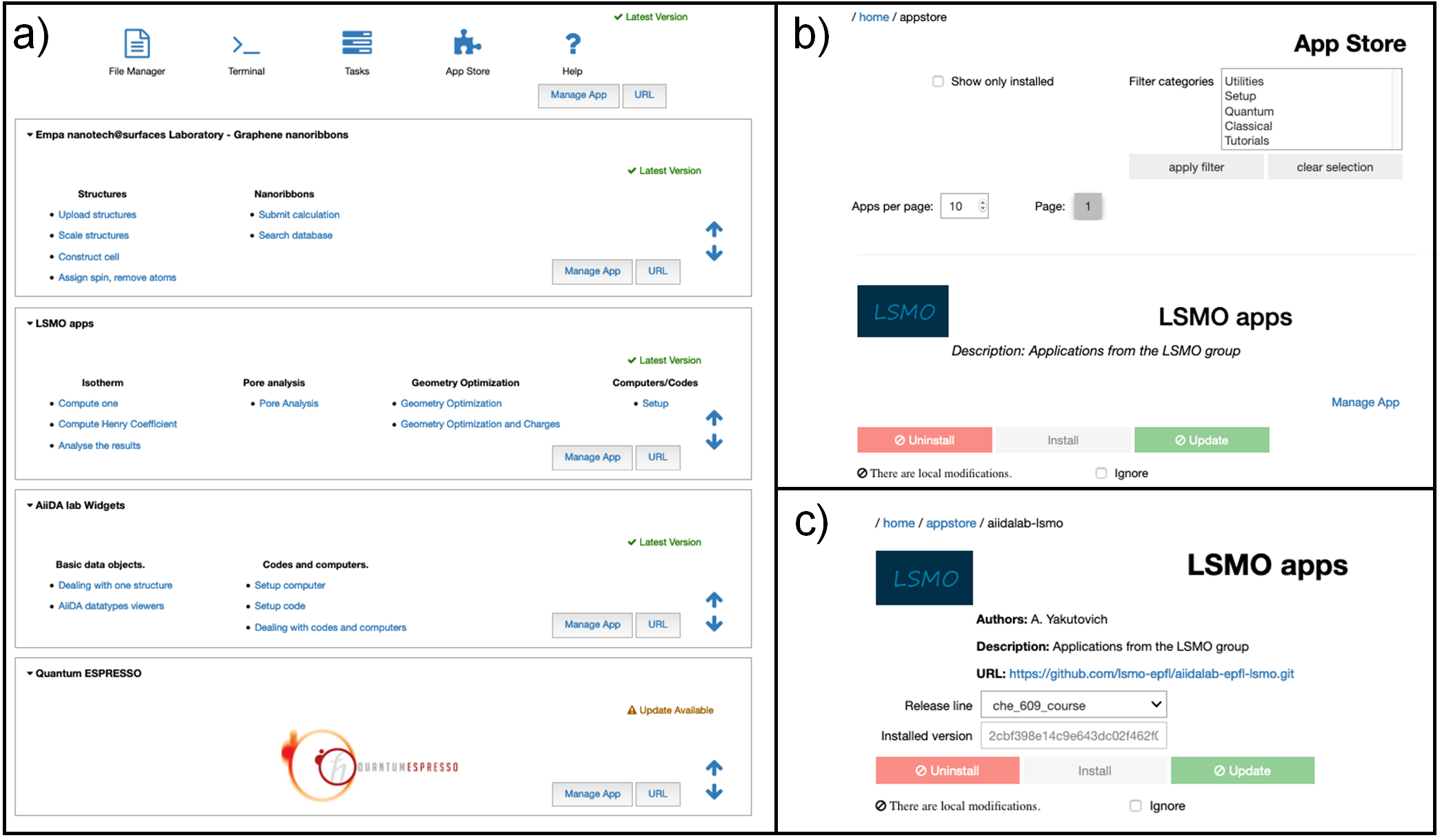}
\caption{(a) Snapshot of the AiiDAlab starting page. At the top, the links to five home applications are visible. Below, additional panels are present for each of the apps installed via the App Store. The viewing of this list can be customized, and in this example includes (among others) the Nanoribbon app discussed in Section \ref{nanoribbons}. (b) Snapshot of the App Store application. The user can select the number of apps shown per page, filter by application groups, and choose whether to show all apps or only the installed ones. (c) Snapshot of the App management page that provides more information about the chosen app and allows selection of a specific version.}
\label{home_app}
\end{figure}

The \textbf{AiiDAlab home app}~\cite{noauthor_aiidalab_2019-2} is the core application of the service.
It takes care of displaying the Home page (Fig.~\ref{home_app}a) and enables basic interactions with the AiiDAlab machine.
It contains five icons for quick access to handy tools: \textit{File Manager} to browse files,  \textit{Terminal} to access a Linux terminal directly from the browser, \textit{Tasks} to list all running AiiDAlab apps, \textit{App Store} to install new applications and \textit{Help} referencing to the AiiDAlab documentation.
\textit{App Store} is a user-friendly interface to the Git repositories of the available apps, which are listed on the AiiDAlab registry (see Section \ref{subsection_registry}); it allows to install, uninstall, and update apps with just one click (Fig.~\ref{home_app}b).
When installing an app, the user can choose between its various versions defined by the author in the corresponding Git repository (Fig.~\ref{home_app}c).

\subsection{Publishing AiiDAlab applications}
\label{subsection_registry}
Available AiiDAlab apps are listed in a central ``AiiDAlab registry''~\cite{noauthor_aiidalab_2019}. Once an app is added to the registry, it is automatically listed in the AiiDAlab App Store of every AiiDAlab user. In order for this to work, the app must be in a Git~\cite{noauthor_git_nodate} version-control system repository accessible by AiiDAlab, like e.g. GitHub or GitLab~\cite{noauthor_github_nodate,noauthor_gitlab_nodate}.
Each entry in the registry contains some minimum information about an app: its name, a link to its Git repository, a link to a metadata file, and a list of categories it falls within. 
Such an infrastructure provides AiiDAlab developers an efficient way to share their scientific expertise and knowledge (encoded into the application and the workflows) with the rest of the community, while maintaining full ownership and control on the development of the apps.
In addition, AiiDAlab allows to share simulation data with their whole provenance (i.e. the history of how each item has been generated, by which calculation and with which input) thanks to the use of AiiDA as the underlying workflow engine.


\subsection{Distribution and availability of AiiDAlab}

We provide two deployment solutions for AiiDAlab: a ``distributed model'' and a ``centralized model'', that are described below.
From the interface point of view, both distributed and centralized AiiDAlab solutions are identical.

The distributed model follows the traditional way of sharing software. To share an entire stack, we provide a Virtual Machine (VM) image for download, which users can run locally. This is the case of Quantum Mobile~\cite{noauthor_quantum_nodate} -- a VirtualBox~\cite{noauthor_oracle_nodate} image that includes AiiDAlab pre-configured. The advantage of this approach is that users retain full control of their own data, and  can run on their computers without the need of a central infrastructure or the need of creating login credentials in an online system.
The downside is that users have to take responsibility for most of the administrative work, such as backup and updates.
Another disadvantage of the distributed model is that there is no control on which VM image version a specific user will be working on, which requires additional work to make sure that apps remain portable. 

The centralized model is generally referred to as Software-as-a-Service (SaaS) or Platform-as-a-Service (PaaS).
To date we provide the following centralized AiiDAlab platforms as a service~\cite{noauthor_aiidalab_nodate-3}: one on cloud servers hosted at the Swiss Supercomputing Center (CSCS), and one on the CESNET servers in Czech Republic (with resources granted by EOSC-hub)\footnote{The EOSC-hub instance is connected to European Grid Infrastructure (EGI) check-in for authentication, giving access to a wide community of researchers.}, both providing high-scalability features via Kubernetes.
This approach can be replicated elsewhere straightforwardly.
The centralized model lowers the entrance barrier for a user, since only login credentials and a web browser are required to access the cloud service, without the need to install custom software. 
At the same time, however, this model requires web engineers to administer and maintain the service.
In order to work in Kubernetes~\cite{noauthor_production-grade_nodate}, AiiDAlab is distributed also within a docker container.
We emphasise that the same AiiDAlab docker container which runs on Kubernetes can also be run locally if the users wants to.
The necessary setup scripts are provided in the corresponding repository~\cite{noauthor_docker_2019}. More details about the AiiDAlab docker container are provided in the ``AiiDAlab docker stack and AiiDAlab package'' section of the Supplementary Information.

Last, it should be noted that companies could have strict policies on the physical location of their data, often requiring ``in house'' storage capabilities.
For this reason, we have decided to license all core parts of AiiDAlab with the industry-friendly MIT open source licence~\cite{noauthor_mit_nodate} allowing for companies to deploy a custom AiiDAlab server on their premises for internal use.
Moreover, to facilitate re-deployment and setting up a (new) AiiDAlab instance, we provide automated scripts to re-build the whole infrastructure using the Ansible tool~\cite{hat_ansible_nodate}.
Specifically, two Ansible roles are provided: ansible-role-aiidalab~\cite{noauthor_marvel-nccransible-role-aiidalab_2019} that installs the AiiDAlab environment on Linux-like operating systems, and ansible-role-aiidalab-server~\cite{noauthor_aiidalabansible-role-aiidalab-server_2020} that deploys a multi-user AiiDAlab server on Ubuntu.
Additionally, to set up an automatically scaling platform, we provide step-by-step installation instructions~\cite{noauthor_aiidalab_nodate} to deploy AiiDAlab on a Kubernetes~\cite{noauthor_production-grade_nodate} cluster.

\section{AiiDAlab use cases}
\label{app_examples}
In order to showcase the benefits of the AiiDAlab platform, we describe some use cases in which it has already been used in production, driven by the needs of experimental research in on-surface chemistry. This is a powerful method toward bottom-up fabrication of complex nanostructures, unattainable via traditional solution chemistry~\cite{gourdon_-surface_2016}.
In this field, the nanotech@surfaces laboratory at Empa combines scanning probe microscopy techniques and atomistic simulations to study polymerization of molecules on catalytic surfaces.
In 2010 the laboratory demonstrated a bottom-up approach~\cite{Cai2010} for the synthesis of graphene nanoribbons (GNRs) and since then a number of GNRs with different topologies~\cite{talirz_-surface_2016} have been fabricated with atomic precision. 
To understand the fabrication process and to predict the electronic properties of the nanomaterials, state-of-the-art computational methods have to be applied routinely with protocols that have been optimized over the last ten years. 
A bottleneck in the discovery process then becomes the ``human factor'': the time needed to manually set up calculations, process them, and identify and correct human errors. 
The classes of calculations that the laboratory deals with include electronic-structure characterization of gas-phase GNRs, identification of intermediate states and activation barriers for the chemical reactions involving molecular precursors adsorbed on noble metal substrates, and on-surface scanning probe microscopy and spectroscopy.

Our goal then has been to create AiiDAlab applications with intuitive user interfaces to perform automated workflows for the calculations mentioned above. We designed the apps in such a way so that they could  be used also by researchers who are not familiar with the details needed to set up a calculation.

\subsection{Nanoribbon app}
\label{nanoribbons}
Obtaining a specific GNR structure requires a considerable effort from the experimental point of view, both in terms of solution chemistry processes needed  to fabricate the appropriate molecular precursors and in terms of the ``on-surface synthesis'' itself. 
Key steps before entering the experimental effort are the prediction of the relevant electronic properties associated to GNRs with a specific geometry and understanding how their topology and composition (for example insertion of heteroatoms) allow to tune such properties.
The Nanoribbon AiiDAlab app~\cite{noauthor_empa_2020-2} has been designed, in direct collaboration with experimentalists, to streamline these tasks.

\begin{figure}[H]
\centering\includegraphics[width=1.0\linewidth]{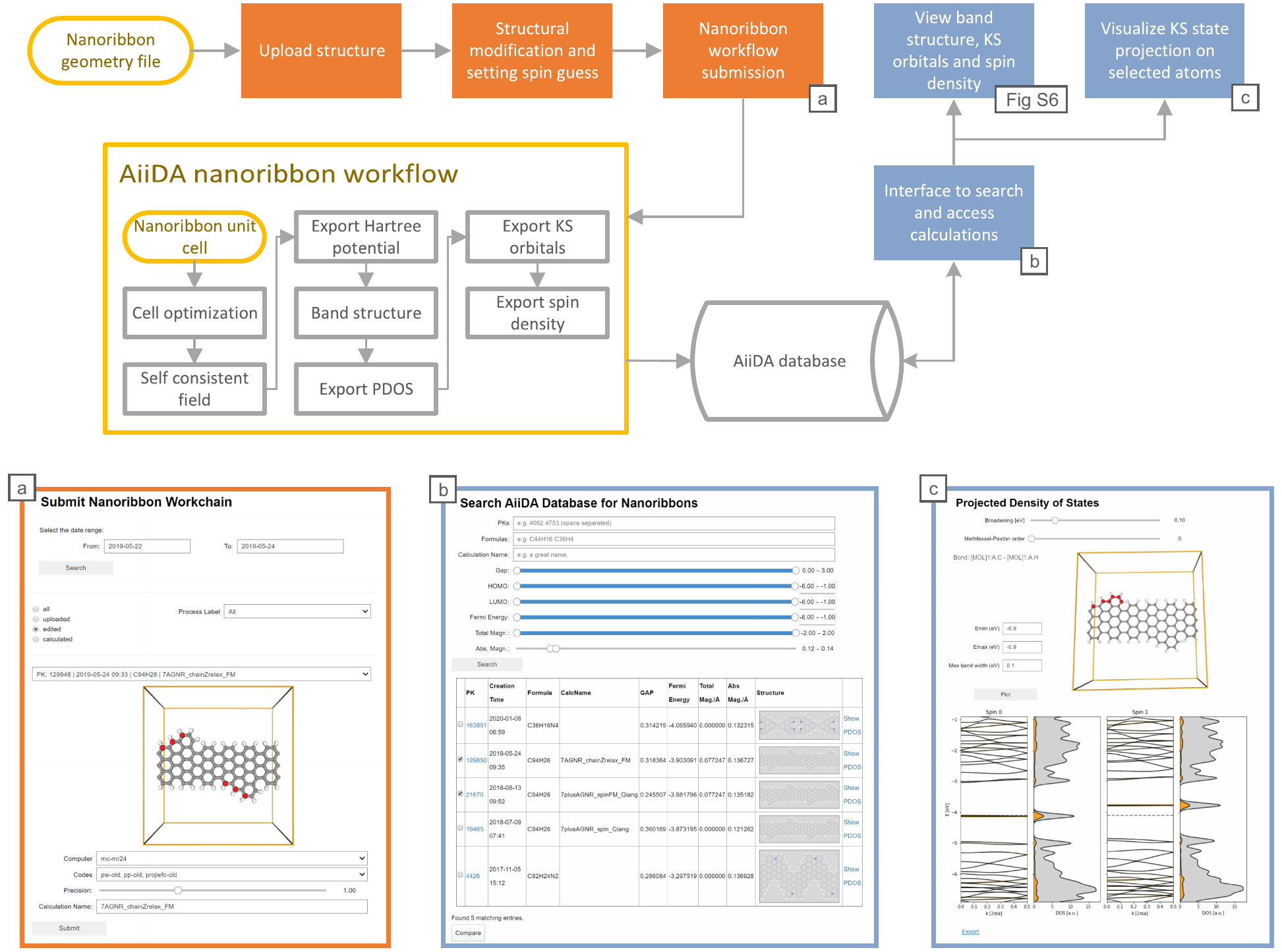}
\caption{Conceptual flowchart of the Nanoribbon app together with three screenshots of user interfaces, shown in panels (a), (b) and (c). Oval shapes represent starting points and filled rectangles represent Jupyter user interfaces for the various steps of preparation (orange) and results (blue). A selection of the user interfaces can be seen in the figures/panels referenced in the bottom right corner of the corresponding rectangle. Arrows highlight logical precedence, where the output of a preceding step becomes the input to the next one, together with additional manual input, if necessary. The AiiDA Nanoribbon workflow sub-chart shows an overview of the calculation steps that the AiiDA daemon subsequently submits to the computational resource. Each step is stored in the AiiDA database.}
\label{fig:nanoribbon_flowchart}
\end{figure}
The app provides a user friendly interface to run the automatic characterization of one-dimensional nanostructures using the Quantum ESPRESSO distribution~\cite{giannozzi_quantum_2009}. The calculation steps involve cell and geometry optimizations, band structure calculations, and exports of various electronic structure properties. Overview of the various user interfaces and calculation steps is shown as a flowchart in Fig.~\ref{fig:nanoribbon_flowchart}. Additional details about the Nanoribbon app can be found in the relevant section of the Supplementary Information.

\subsection{On-Surface Chemistry app}
\label{reactions_app}
To deal efficiently with the computational tasks needed to interpret and predict the outcome of experiments related to on-surface chemistry, the On-Surface Chemistry AiiDAlab app~\cite{noauthor_empa_2020} allows to run a variety of first-principles calculations at multiple levels of theory, using the CP2K simulation package~\cite{hutter_cp2k:_2014}.

\begin{figure}[!ht]
\centering\includegraphics[width=1.0\linewidth]{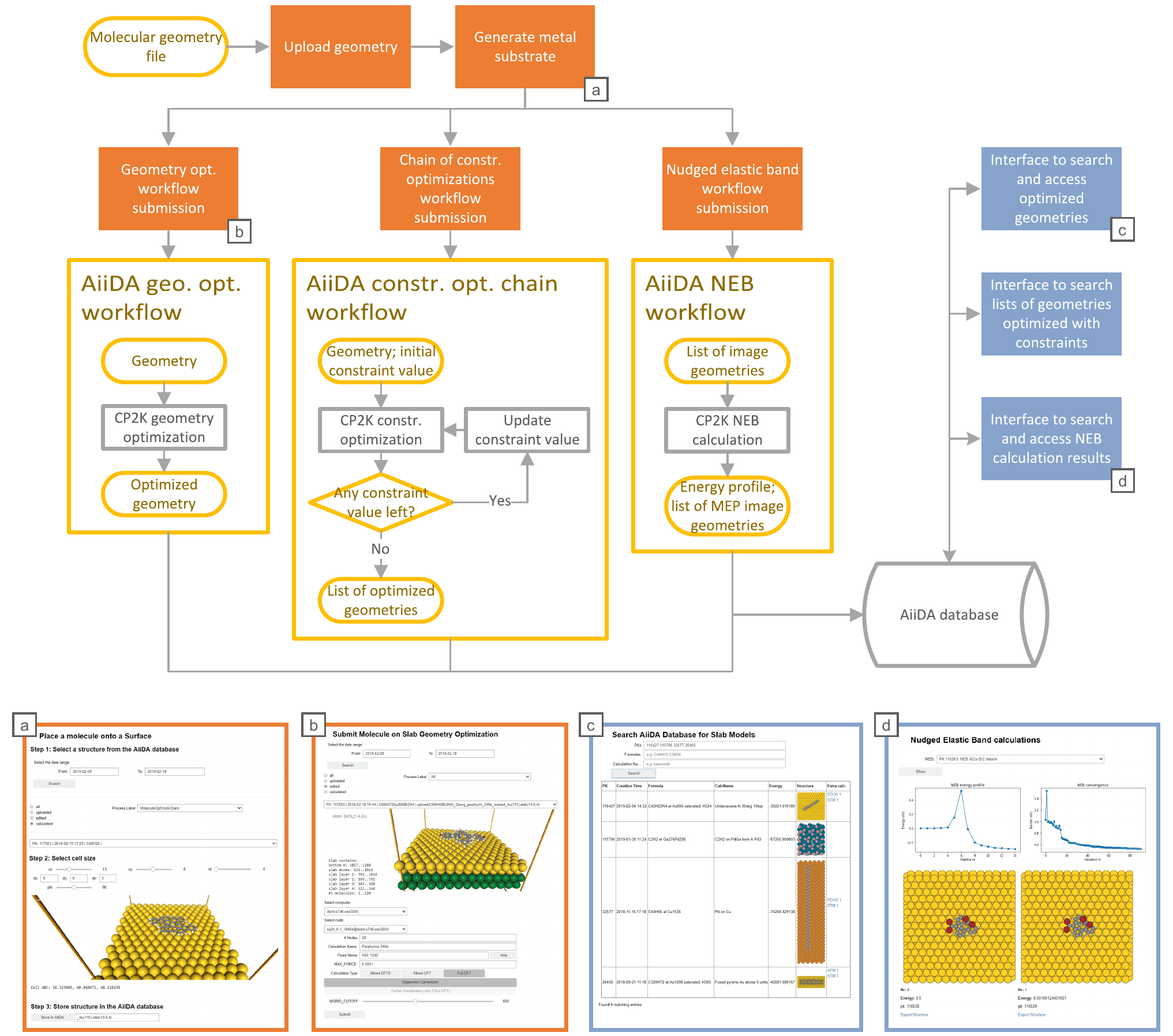}
\caption{Conceptual flowchart of the On-Surface Chemistry app together with four screenshots of user interfaces, shown in panels (a), (b), (c) and (d). See the caption of Fig.~\ref{fig:nanoribbon_flowchart} for the meaning of the various flowchart elements.}
\label{fig:chemistry_flowchart}
\end{figure}

The app provides workflows to run geometry optimizations and nudged-elastic-band (NEB) calculations \cite{mills_quantum_1994,henkelman_climbing_2000}. Additionally, a workflow is provided to compute a physically feasible initial path for a NEB calculation using a chain of constrained geometry optimizations: Starting from an optimized geometry, the user defines a collective variable (CV) (e.g. a bond length) and a list of values that the CV should fulfill in sequence. After the geometry of the current step reaches equilibrium, fulfilling the current constraint for the CV, a new optimization starts imposing the next constraint value in the list. This procedure allows to identify a reasonable guess for the reaction path when linear interpolation between an initial and a final state would result in a non-physical path. Several supporting tools are provided as Jupyter notebooks, allowing for instance to modify structural geometry, create metal slabs, and conveniently access and export the calculation results directly from the browser. An overview of the various user interfaces and calculation steps is shown as a flowchart in Fig.~\ref{fig:chemistry_flowchart}. Additional details about the On-Surface Chemistry app can be found in the Supplementary Information.

\subsection{Scanning Probe Microscopy app}
\label{spm_app}

\begin{figure}[!ht]
\centering\includegraphics[width=0.8\linewidth]{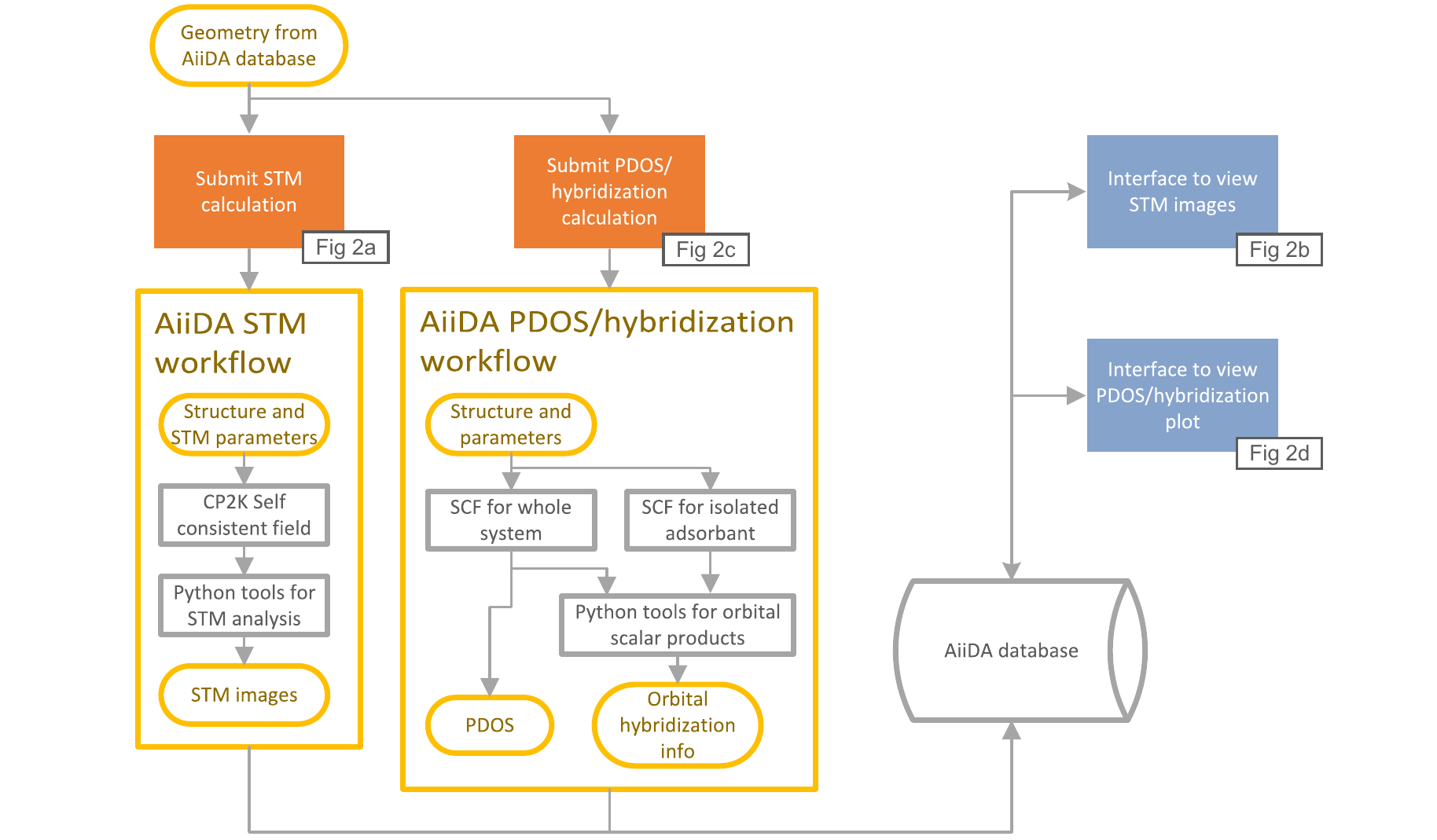}
\caption{Conceptual flowchart of the Scanning Probe Microscopy app. See the caption of Fig.~\ref{fig:nanoribbon_flowchart} for the meaning of the various flowchart elements.}
\label{fig:spm_flowchart}
\end{figure}

Scanning probe microscopy is the main experimental tool to acquire structural and electronic information about molecules and nanostructures on metal surfaces. 
More specifically, scanning tunnelling microscopy (STM) and spectroscopy (STS) allow to determine the spatial distributions of the electronic states and the electronic band gap of the system under investigation.
However, to validate the correctness of the assignment of the electronic states, to investigate the effect of the substrate, and to determine the geometry of the system, experimental electronic signatures need to be compared with computational simulations.
To run these simulations, we have developed the Scanning Probe Microscopy app~\cite{noauthor_empa_2020-1}.
The app includes a workflow for the simulation of STM/STS images and a workflow to run projected density of states (PDOS) analysis together with orbital hybridization investigation for molecules adsorbed on metal slabs. An overview of the various user interfaces and calculation steps is shown as a flowchart in Fig.~\ref{fig:spm_flowchart}.

The interfaces for submission and for analysis of the result for the two workflows provided in the Scanning Probe Microscopy app as shown in Fig.~\ref{fig:spm}. In the submission interface of the STM/STS workflow (Fig.~\ref{fig:spm}a), the user can specify the input geometry from the AiiDA database, the remote computer and codes, and the STM parameters. The corresponding viewing interface, shown in Fig.~\ref{fig:spm}b, allows to visualize a series of images by specifying discrete bias voltage values or a continuous range of energies. In the example shown, a discrete series of images are visualized for a porous nanographene molecule. Additionally, the selected visualizations can be exported as a ZIP file archive containing the images in PNG, IGOR Pro and raw data formats. The orbital hybridization workflow submission interface contains the same geometry and code selection widgets as the STM/STS workflow with the additional widgets illustrated in Fig.~\ref{fig:spm}c. The corresponding viewing interface, shown in Fig.~\ref{fig:spm}d, allows to inspect the PDOS and the orbital overlaps with an effective interactive plot that combines the two. Further details about the Scanning Probe Microscopy app are presented in the Supplementary Information.

\section{Conclusions and Outlook}
\label{conclusion}
We presented the overall concept and the implementation of AiiDAlab, an ecosystem for the development, sharing and deployment of user-friendly simulation apps that 1) provides convenient user interfaces to computational workflows, hiding their technical details and making them straightforwardly available to a broad range of researches, and 2) facilitates the development of such interfaces via a Python-based programming language and the availability of common interactive widgets. 

AiiDAlab offers the possibility to contribute and share computational workflows and data: the software stack is standardized using Docker containers, which provide every user with the same identical software environment.
In order to minimize the entrance barrier for new users, the platform can run directly ``in the cloud'', requiring only a web browser and a login to access it (servers are currently hosted at CSCS, Switzerland and  EOSC, Czech Republic), but can also be redeployed on internal resources or run locally using the Quantum Mobile virtual machine~\cite{noauthor_quantum_nodate}.
AiiDAlab leverages the flexibility of the Python language and the power of Jupyter Notebooks to expose a simple programming interface to app developers.
Additionally, in order to provide an intuitive user interface, a novel Jupyter extension called Appmode has been developed.
Together with Jupyter and the use of widgets (both existing ones, and new ones developed by us), these tools make it possible to quickly create web applications in Python, taking advantage of existing libraries like ipywidgets~\cite{noauthor_jupyter-widgets/ipywidgets_2019}, matplotlib~\cite{noauthor_matplotlib_nodate}, bqplot~\cite{noauthor_bqplotbqplot_2020}, nglview~\cite{noauthor_nglviewernglview_2020}, and ase~\cite{larsen_atomic_2017}. Since Python knowledge is widespread in the community, we expect that many scientists will be able to contribute new apps.

We also provide examples of a full range of apps which were developed in collaboration with experimentalists to tackle specific problems in the field of carbon-based nanostructures and surface-supported chemical reactions, enabling to run complex simulations effortlessly. 

In addition to the applications presented here, that demonstrate the benefits and functionality of the platform, numerous more apps are currently under development. To facilitate this process, we provide a set of tools for the creation of new apps, including a ``cookie cutter'' recipe that automates the initial steps of the app creation and the AiiDAlab widgets which can be readily embedded into new apps (see Supplementary Information). 

Last, in order to encourage not only the sharing of data, but also the sharing of codes, workflows and applications within the community, we deployed an open AiiDAlab registry. Developers can use it to register their apps and make them automatically available to all other users via the App Store. 

AiiDAlab is a modular and efficient platform granting a collaborative environment centered on simulations and targeted to people with different expertise. The platform can serve a broad range of communities beyond the original focus on materials science, and facilitates sharing of turn-key computational workflows and of the data produced by simulations.
AiiDAlab allows for a direct exchange of processed simulation results in an appropriate format, introduces an easy way to standardize routine calculations, reducing to a minimum bottlenecks such as human mistakes.
We expect that computational scientists from diverse fields will start adopting the platform and contributing to it with workflows and interfaces, empowering the scientific community by making simulation tools more directly and easily accessible to a broader range of researchers.

\section*{Acknowledgements}
This work is supported by the MARVEL National Centre for Competency in Research funded by the Swiss National Science Foundation (grant agreement ID~51NF40-182892), the European Centre of Excellence MaX ``Materials design at the Exascale'' (grant no.~824143), 
the ``MaGic'' project of the European Research Council (grant agreement ID~666983),
the swissuniversities P-5 ``Materials Cloud'' project (grant agreement ID~182-008),
the ``MARKETPLACE'' H2020 project (grant agreement ID~760173),
the ``INTERSECT'' H2020 project  (grant agreement ID~814487),
the ``EMMC'' H2020 project (grant agreement ID~723867), the ``OSSCAR'' project (funded by the EPFL Open Science Fund), and the Swiss National Science Foundation grant agreement ID~172527.
We acknowledge PRACE for awarding us simulation time on Piz Daint at CSCS (project ID~2016153543) and Marconi at CINECA (project ID~2016163963), and the support of Swiss Platform for Advanced Scientific Computing PASC.

We thank the CSCS and CESNET support teams for continued support during the deployment of AiiDAlab on their infrastructures, and the EOSC-Hub Early Adopter Programme for providing kubernetes resources.

\section*{Data Availability}
All the code of the AiiDAlab platform as well as the AiiDAlab applications discussed in the paper are open source (released under the MIT license) and can be downloaded from the links provided in the References section.

\section*{References}
\bibliographystyle{bib-style}
\bibliography{aiidalab}
\end{document}


\begin{frontmatter}


\title{Supplementary information: AiiDAlab -- an ecosystem to develop, execute and share scientific workflows}

\author[marvel,theos,lsmo]{Aliaksandr V. Yakutovich \textsuperscript{*,}\corref{contrib}}
\author[marvel,empa]{Kristjan Eimre\corref{contrib}}
\author[marvel,empa]{Ole Schütt\corref{contrib}}
\author[marvel,theos,lsmo]{Leopold Talirz}
\author[theos]{Carl S. Adorf}
\author[theos]{Casper W. Andersen}
\author[marvel,empa]{Edward Ditler}
\author[marvel,theos]{Dou Du}
\author[marvel,empa]{Daniele Passerone}
\author[marvel,lsmo]{Berend Smit}
\author[marvel,theos]{Nicola Marzari}
\author[marvel,theos]{Giovanni Pizzi\corref{corauthor}}
\author[marvel,empa]{Carlo A. Pignedoli\corref{corauthor}}

\address[marvel]{National Centre for Computational Design and Discovery
of Novel Materials (MARVEL), \'Ecole Polytechnique F\'ed\'erale de Lausanne, 
CH-1015 Lausanne, Switzerland}
\address[theos]{Theory and Simulation of Materials (THEOS), 
    Facult\'e des Sciences et Techniques de l'Ing\'enieur, 
    \'Ecole Polytechnique F\'ed\'erale de Lausanne,
    CH-1015 Lausanne, Switzerland}
\address[lsmo]{Laboratory of Molecular Simulation (LSMO),
    Institut des Sciences et Ingenierie Chimiques,
    Valais, \'Ecole Polytechnique F\'ed\'erale de Lausanne,
    CH-1951 Sion, Switzerland}
\address[empa]{nanotech@surfaces laboratory,
    Swiss Federal Laboratories for Materials Science and Technology (Empa),
    CH-8600 D\"ubendorf, Switzerland}

\cortext[corauthor]{Corresponding author}
\cortext[contrib]{Authors contributed equally}


\begin{keyword}
Simulations \sep High-throughput \sep Materials database \sep Scientific workflows \sep Web based platform  \sep Provenance \sep Data management


\end{keyword}

\end{frontmatter}

\section{AiiDAlab application}

\begin{figure}[H]
\centering\includegraphics[width=1.0\linewidth]{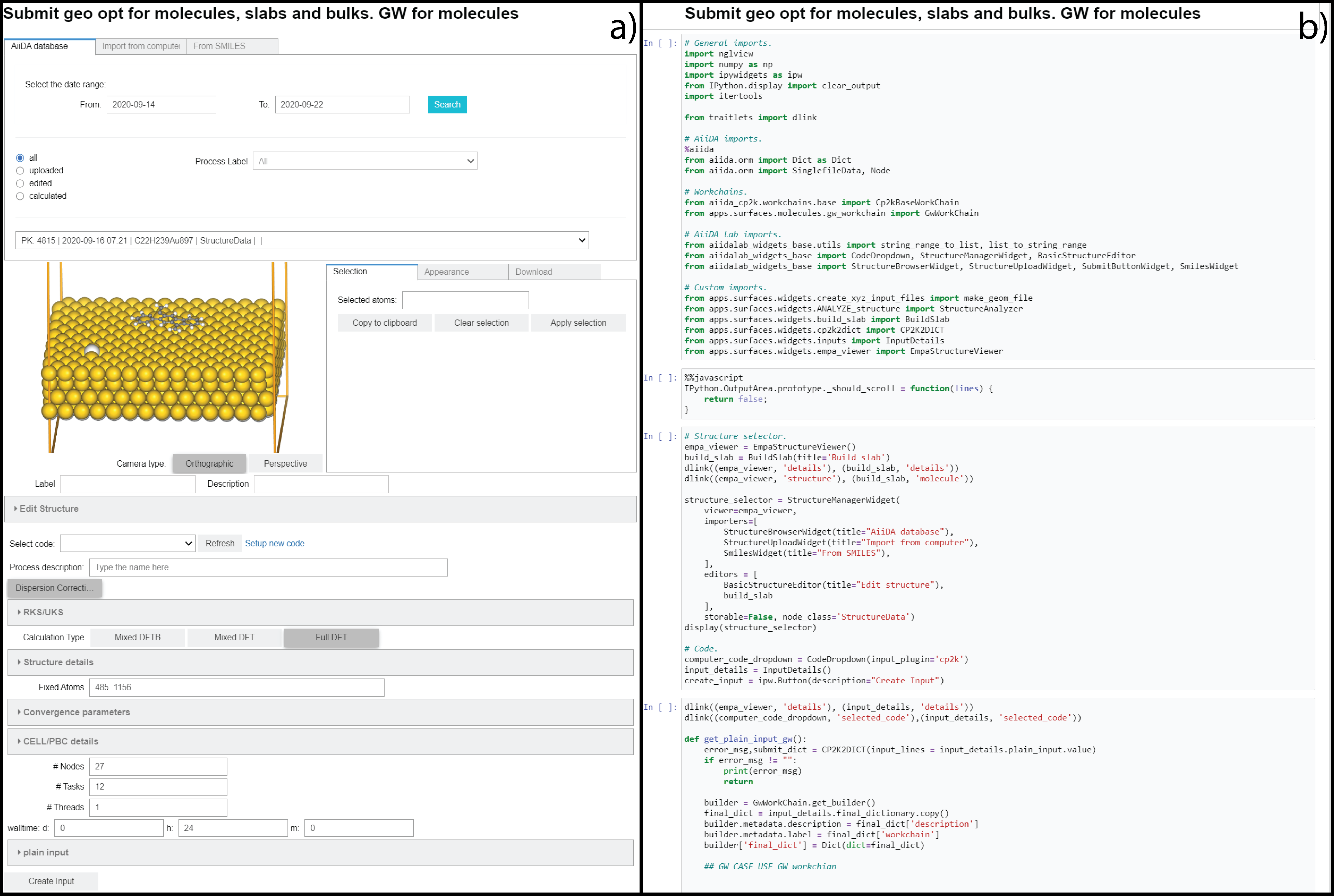}
\caption{Example of an AiiDAlab application that allows to optimize the geometry of a molecule-slab complex. The quantum engine used for optimization is CP2K. Panel a) shows the application interface using Appmode, while panel b) shows the corresponding source code.}
\label{fig:aiidalab_app}
\end{figure}

From a technical point of view, an AiiDAlab application is a folder consisting of Jupyter notebook(s), \textit{metadata.json}, and \textit{start.py} (or \textit{start.md}) files.
When opened, jupyter notebooks are converted into user friendly web based interfaces by the Appmode plugin~\cite{schutt_appmode:_2019} that is enabled by default in AiiDAlab. The role of Appmode is to hide all input code cells and automatically execute all of them (see Fig.~\ref{fig:aiidalab_app}).
Only cells' outputs are thus shown, and in combination with the use of widgets like text boxes, sliders, buttons or more advanced visualizers, the result is an intuitive, interactive ``web application''.

In addition, Appmode allows to have independent copies of the same notebook opened simultaneously and to pass information from one notebook to another via URL parameters. To avoid overload of the server, the Appmode Jupyter kernel is automatically shut down once a user closes the corresponding web page.

\begin{figure}[H]
\centering\includegraphics[width=0.9\linewidth]{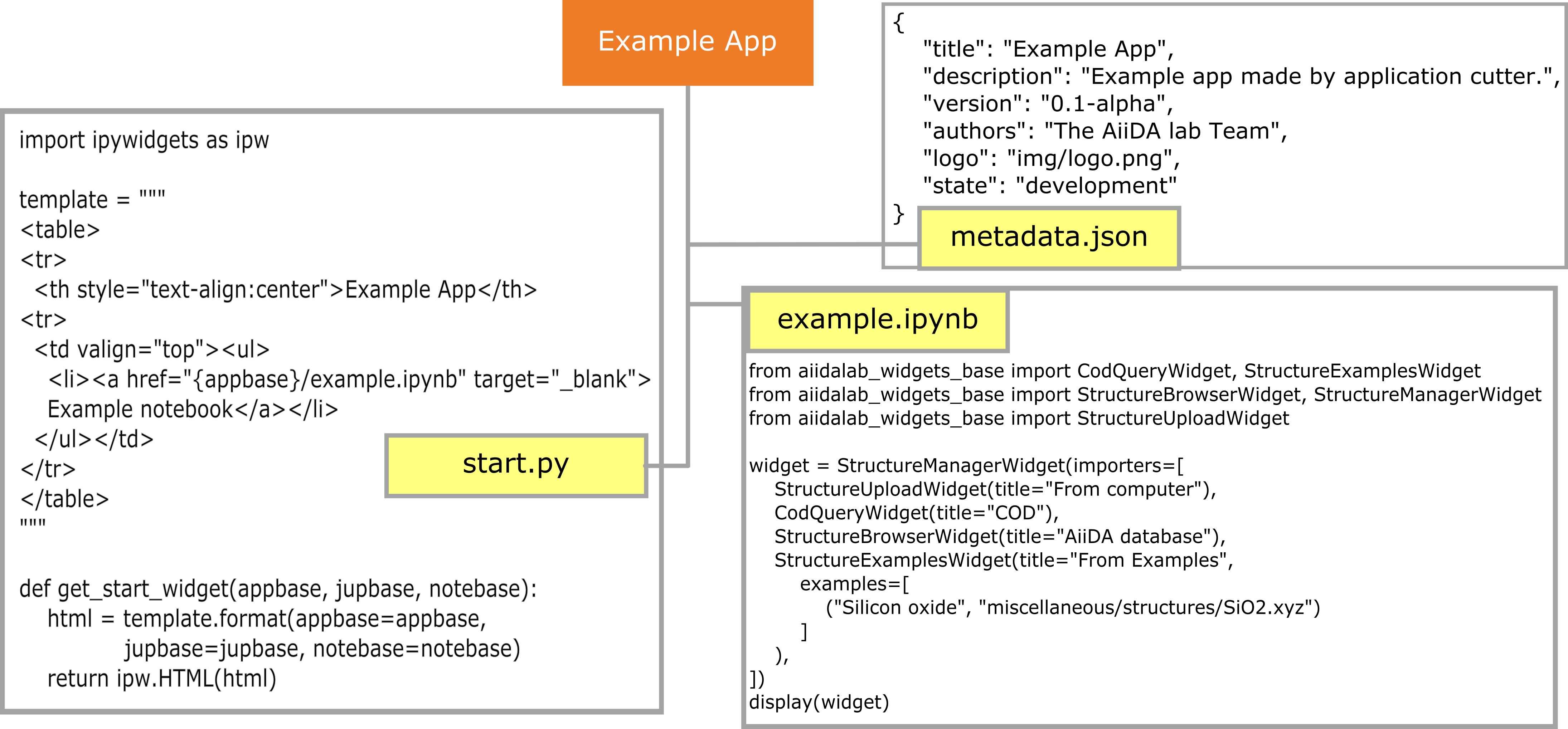}
\caption{Example of a basic AiiDAlab application containing files such as \textit{metadata.json} containing meta-information about the app, \textit{start.py} defining the appearance of the app on the start page, \textit{example.ipynb} jupyter notebook that contains the code for the user interface and calls the underlying code.}
\label{fig:example_app}
\end{figure}

The creation of an AiiDAlab application is a straightforward task.
Any folder created in the ``/home/\{user\}/apps'' directory of the AiiDAlab environment is considered an application and processed accordingly. 
The appearance and behaviour of the app can be controlled by placing specific files in this directory (see Fig.~\ref{fig:example_app}).
The appearance of an application on the home page is defined in a \textit{start.py} (or \textit{start.md}) file.
Any meta information like the list of authors, a description or a logo can be provided in a \textit{metadata.json} file.
The user interface of the app relies on  a set of accompanying Jupyter notebooks (files with \textit{*.ipynb} extension) which contain the Python code responsible for launching workflows, controlling their execution, displaying and analyzing results, and other programmatic tasks.
AiiDA workflows are usually defined in separate Python files.
To simplify the task of writing an application, avoiding the need to manually create the files mentioned, we provide a dedicated cookie cutter recipe~\cite{noauthor_cookie_2019}, i.e. a template that can be automatically converted into a fully working application by means of the cookiecutter code, thus automating the initial steps of the app creation process.

\section{AiiDAlab docker stack and AiiDAlab package}

When using AiiDAlab in the cloud, every user works in an isolated virtual environment -- a Docker container~\cite{noauthor_docker:_nodate}.
All the necessary information to build the corresponding image is provided by the aiidalab-docker-stack~\cite{noauthor_docker_2019} repository.
The aiidalab-docker-stack image is built on top of the aiida-core image~\cite{noauthor_aiidateamaiida-core_nodate} that is based on aiida-prerequisites~\cite{noauthor_aiidateamaiida-prerequisites_nodate} image.
The base one, aiida-prerequisites, configures the required Linux system users and it sets up the PostgreSQL~\cite{noauthor_postgresql_nodate} and RabbitMQ~\cite{noauthor_messaging_nodate} servers required by AiiDA.
On top of it, the aiida-core docker image adds AiiDA, its Python dependencies and the corresponding configuration scripts. The top level image, aiidalab-docker-stack, includes the remaining required packages e.g. AiiDAlab package~\cite{noauthor_aiidalab_2019-1}, AiiDA plugins, and AiiDAlab home application~\cite{noauthor_aiidalab_2019-2}. It also performs the necessary configuration steps needed to make Jupyter-related components execute properly.

\section{Monitoring submitted calculations}
\label{workflows}
\begin{figure}[H]
\centering\includegraphics[width=0.9\linewidth]{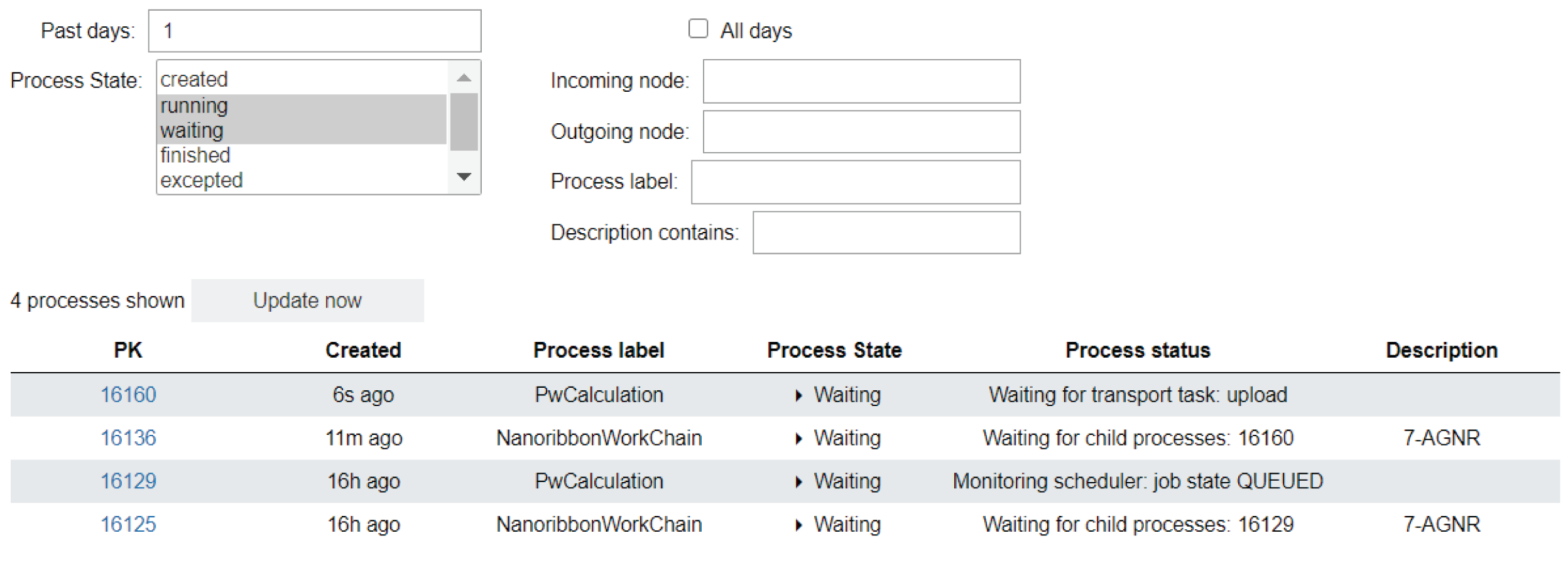}
\caption{This app allows to list all processes that are running, waiting, completed or excepted. Selecting the PK of the submitted workflow allows to follow the different calculations executed (see Fig.~\ref{fig:follow3}).}
\label{fig:follow1}
\end{figure}

AiiDAlab widgets base package~\cite{noauthor_reusable_2018,noauthor_aiidalab_nodate-2} comes with a set of standard tools to monitor the submitted processes.
The ``Process list'' tool allows to monitor the list of processes that were submitted, see Fig.-\ref{fig:follow1}. The user can decide whether to restrict or not the process monitoring to processes belonging to a specific time window. Additionally, the processes can be filtered by the input nodes, output nodes, labels, and words contained in the description. Finally, the user can specify a specific process status to look for: created, running, waiting, finished, excepted. The tool is interactive and its output is updated automatically.

\begin{figure}[H]
\centering\includegraphics[width=0.9\linewidth]{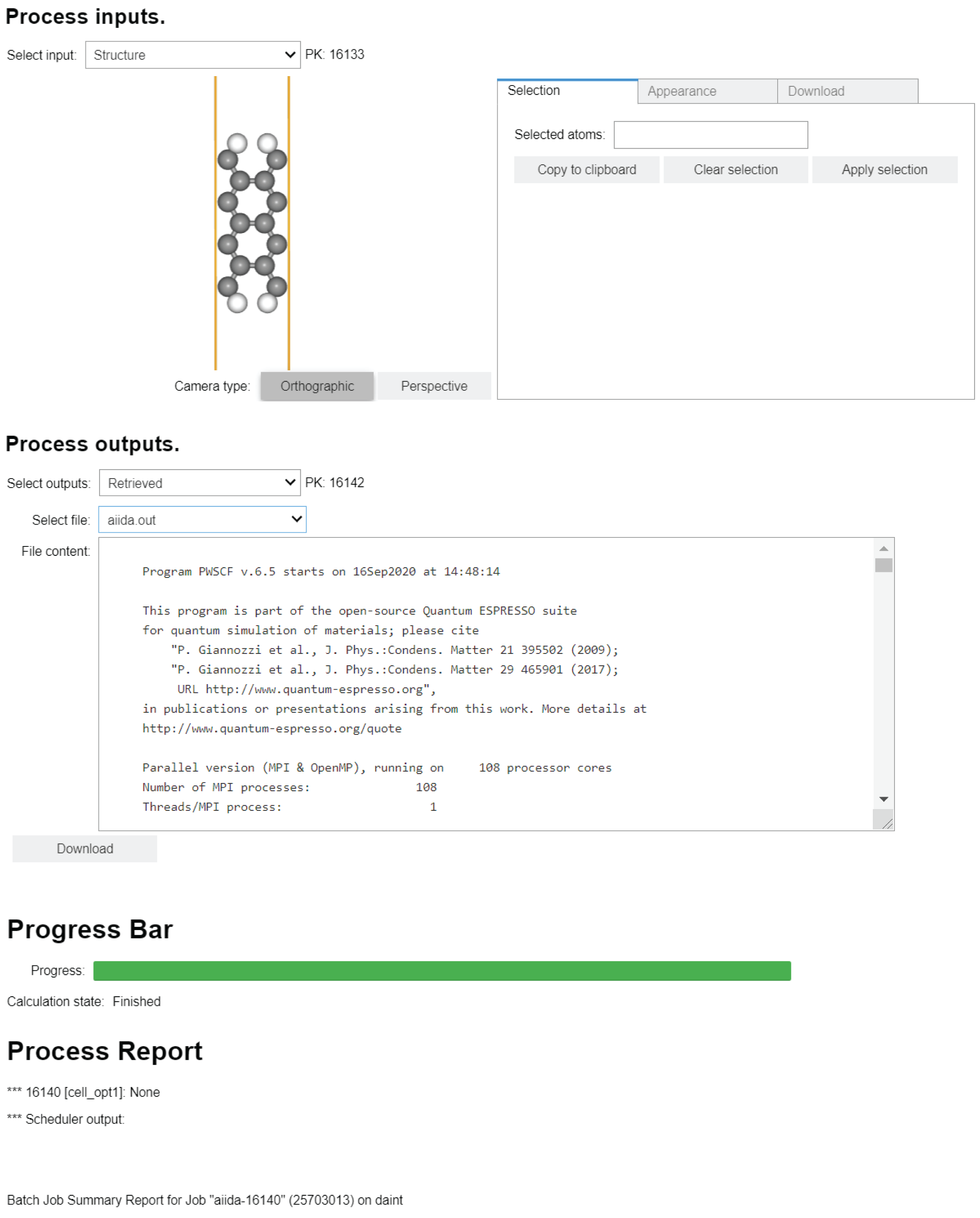}
\caption{The app is showing data such as the input structure, the output produced by the calculation, the scheduler report for one of the calculations in the workflow. }
\label{fig:follow3}
\end{figure}

It is also possible to access a more detailed overview of a specific process.
As shown in  Fig.~\ref{fig:follow3} a user can, for example, visualize inputs and outputs of a process.
Depending on the exact type of  input/output object AiiDAlab will activate a suitable visualiser.
In  Fig.~\ref{fig:follow3}, for example, the input parameter ``Structure'' is visualized using the molecular viewer, while the output file ``aiida.out'' is visualized using a text viewer. In case the selected process is a workflow, its steps (and their status) will be shown as well. The user can select a process (be it a calculation or a workflow) and be redirected to its overview page.

\section{Detailed overview of the Nanoribbon app}
\label{nanoribbons}

\begin{figure}[H]
\centering\includegraphics[width=0.8\linewidth]{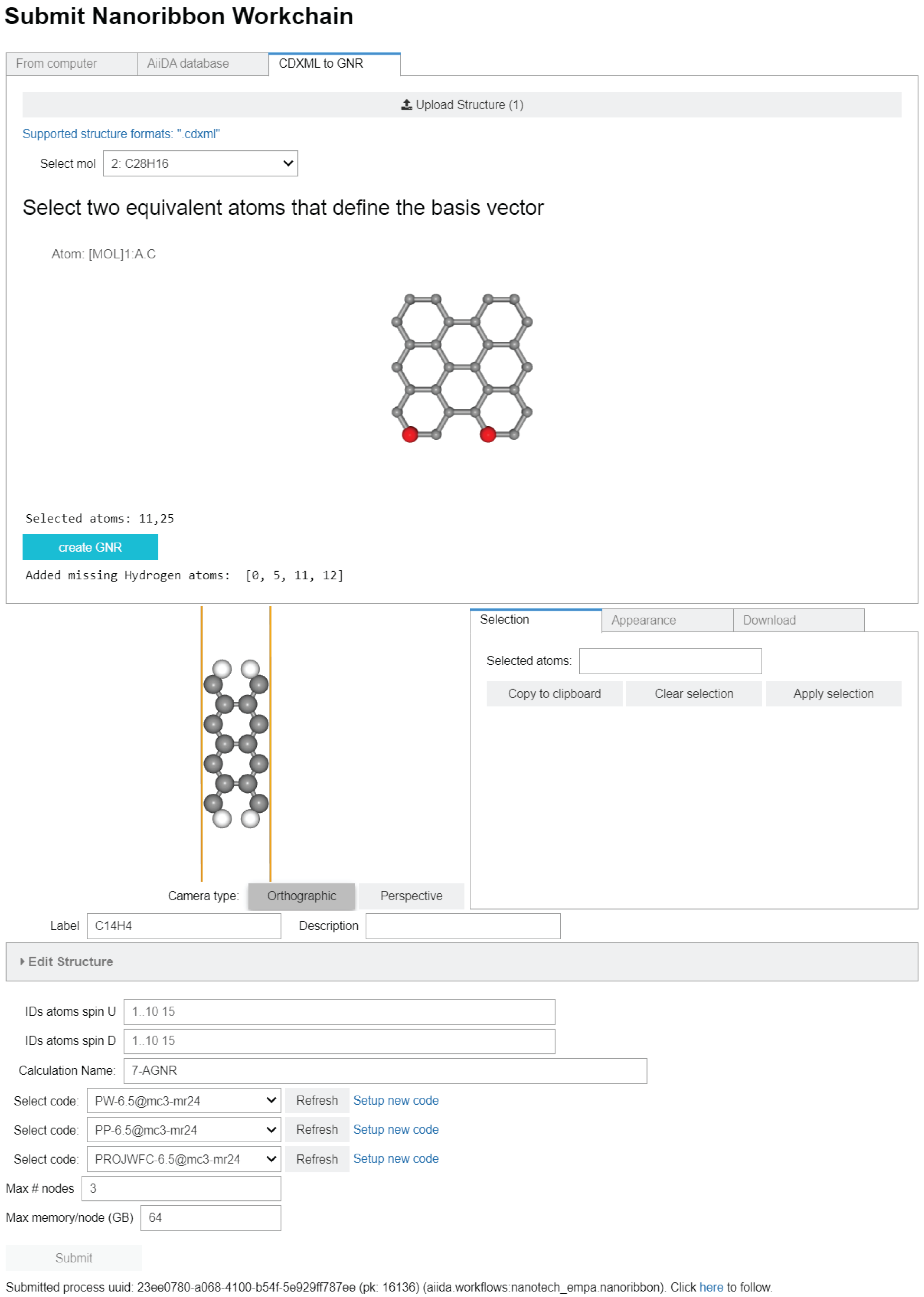}
\caption{A chemical sketch of a ribbon geometry uploaded from a CDXML file. The user selects two equivalent atoms to define the unit cell of the infinite ideal ribbon.
Based on this information the app creates the nanoribbon and adds missing hydrogen atoms.
The structure can be further edited, and atoms can be added and removed.
Once the computational facility where the calculation should be submitted is selected, the workflow can be started right away. It is possible to follow the progress of the workflow clicking directly on the link prompted or using specific apps (see Fig.~\ref{fig:follow1}, \ref{fig:follow3})}
\label{fig:nanoribbons-submit-01}
\end{figure}

The app is designed as follows:
\begin{itemize}
\item A chemical sketch produced with a software for chemistry design is uploaded as a CDXML file. Starting from it, an interactive tool allows the user to define the unit cell of a GNR selecting two equivalent atoms in the structure (Fig.~\ref{fig:nanoribbons-submit-01}). Missing H atoms are automatically added to the edges of the periodic GNR. Alternatively, the correct unit cell of a nanoribbon can be directly uploaded as a standard molecular geometry file or uploaded from the local AiiDA database.
\item An interactive tool allows the user to add or remove atoms from the ribbon and to define an initial guess for the spin distribution.
\item Submission of the workflow will perform a cell optimization, a band structure calculation, a projected density of states calculation and a calculation of Kohn-Sham orbitals. The Quantum ESPRESSO~\cite{giannozzi_quantum_2009,giannozzi_advanced_2017} package is used for the calculations.
\item The results are automatically organized in a database that can be consulted searching for structures with a specific chemical formula, with specific values for the electronic gap or the total magnetization, or with a specific energy position of the top of the valence band and of the bottom of the conduction band.
\item From the list of structures matching the search criteria, the user can select structures for band structure comparison or access a discovery section where an interactive band structure is displayed (Fig.~\ref{fig:show_nanoribbons}a). Selecting a specific band with the cursor activates a slider that shows the KS orbitals at different k-points for the selected band. Another discovery section displays the band structure together with the projected density of states projected on atomic orbitals localized on specific atoms, selected by the user directly on the displayed geometry of the GNR (Fig.~\ref{fig:pdos_nanoribbons}).

\begin{figure}[H]
\centering\includegraphics[width=1.0\linewidth]{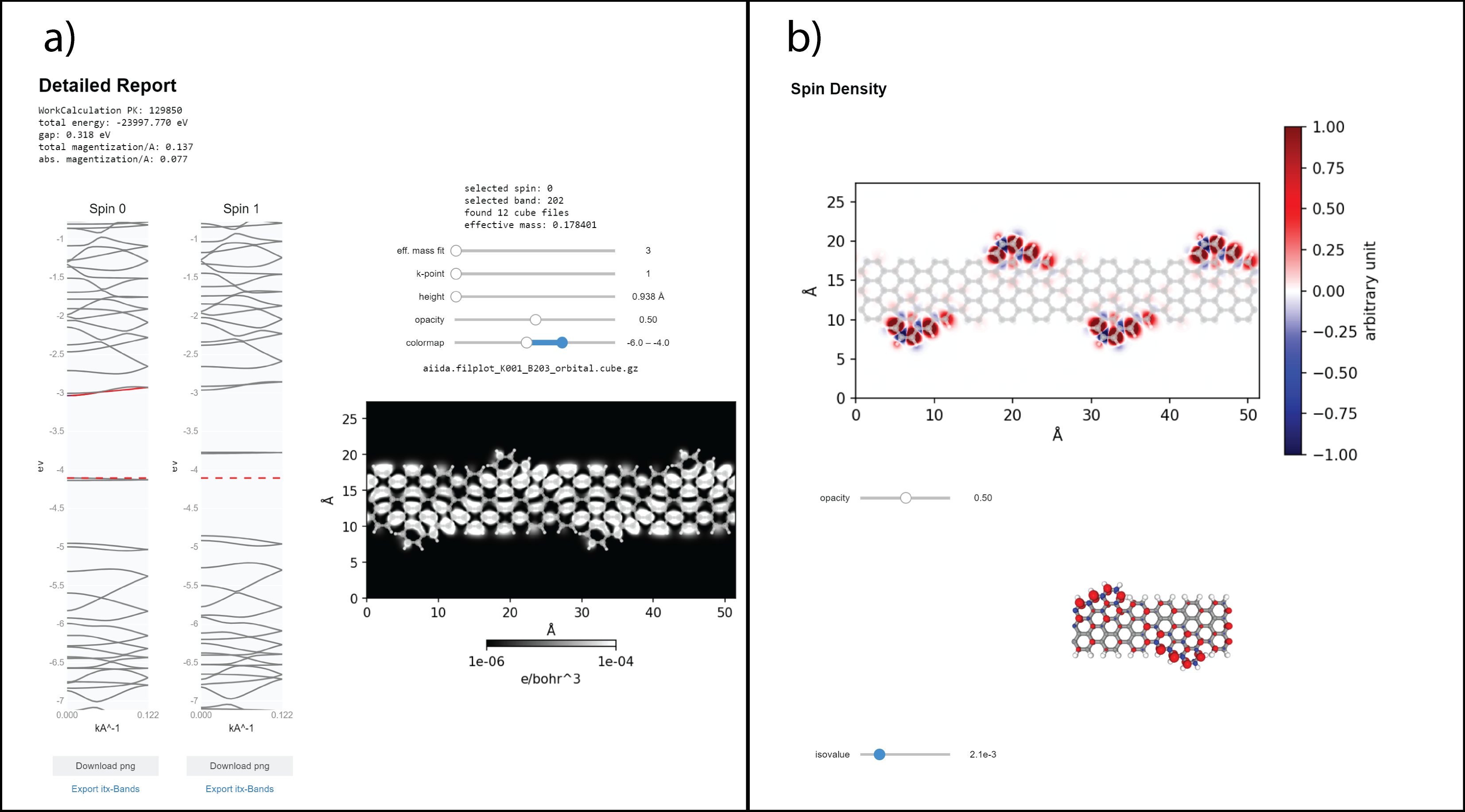}
\caption{Once a specific entry from the nanoribbon database has been identified, it is possible to follow a link to analyze Kohn-Shan orbitals along selected bands (a) and to visualize the absolute spin density (b).}
\label{fig:show_nanoribbons}
\end{figure}

\begin{figure}[H]
\centering\includegraphics[width=1.0\linewidth]{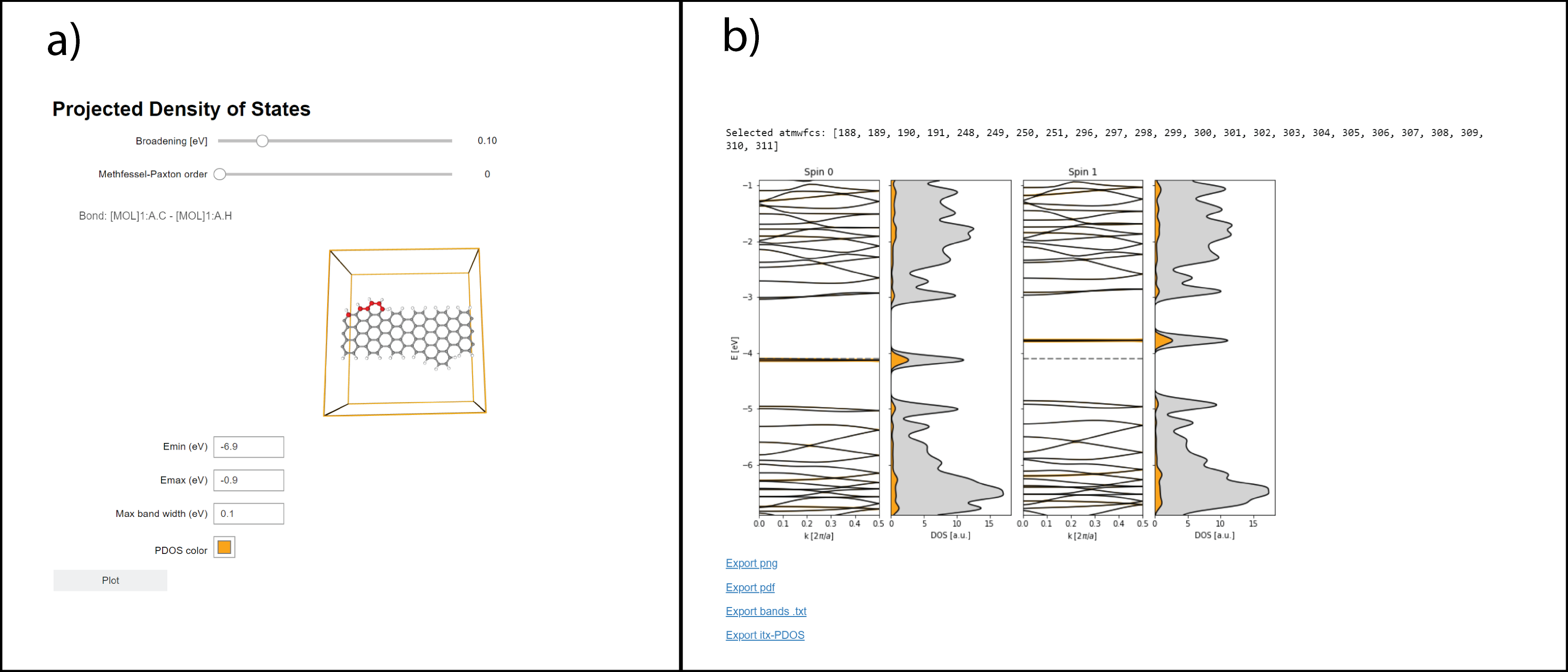}
\caption{Another link in the database of nanoribbons activates the projected-density-of-states visualization interface. Selection of specific atoms (highlighted in red) within the visual GNR model (a) produces the corresponding PDOS plot displayed together with the band structure (b). }
\label{fig:pdos_nanoribbons}
\end{figure}
\end{itemize}

\section{Detailed overview of the On-Surface Chemistry app}
\label{reactions_app}

\begin{figure}[H]
\centering\includegraphics[width=0.8\linewidth]{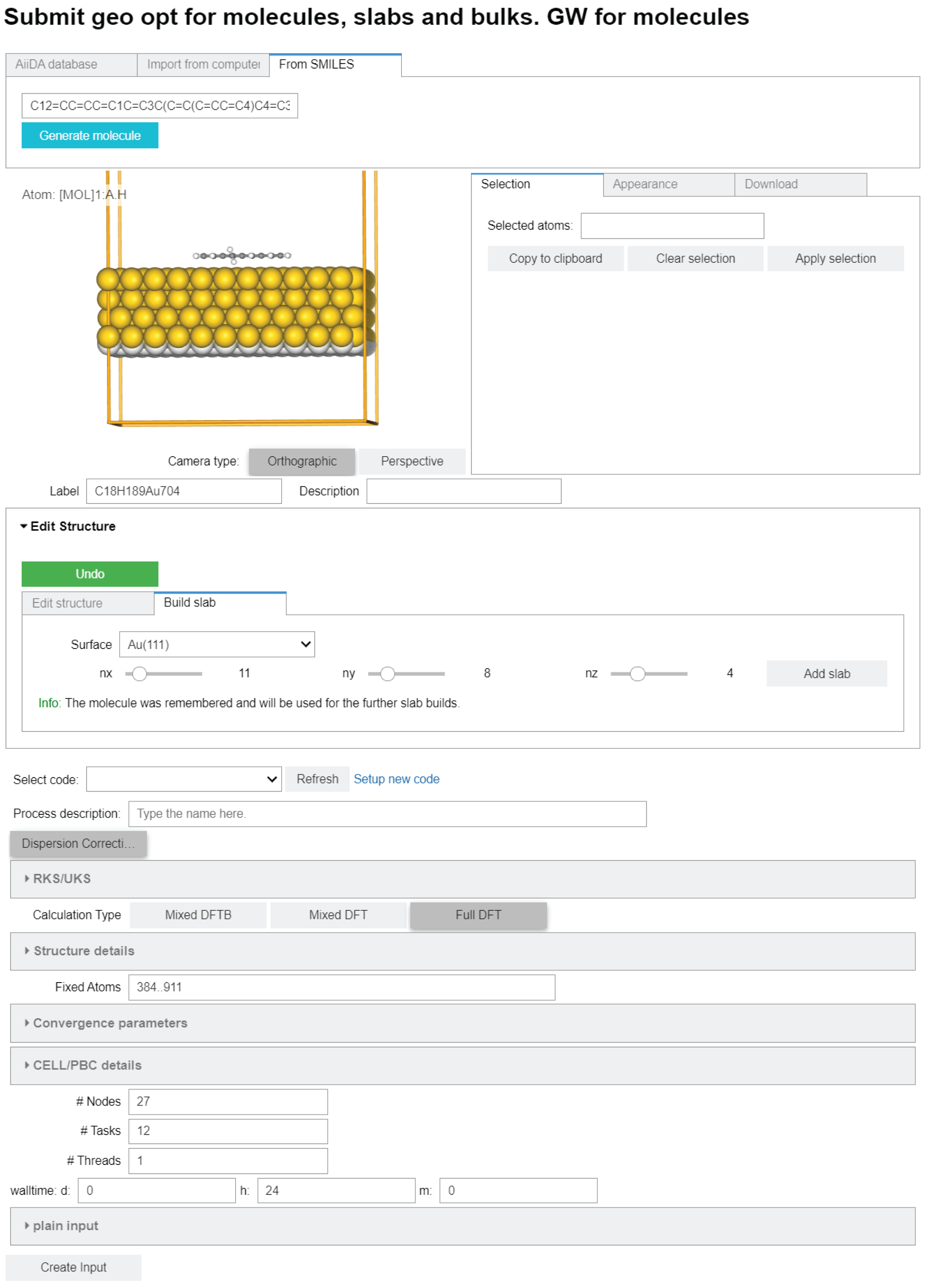}
\caption{The figure shows the web based interface to submit calculations for the optimization of molecules, slabs or bulk systems. In this case we have created a molecule from SMILES, modified its structure and placed the molecule onto a predefined substrate. After that, the geometry is ready to be submitted with parameters preassigned that can be modified accessing the different menus. After clicking the ``Create input'' button, it is possible to inspect the automatically generated input for the calculation. Once this is done, the simulation can be launched.}
\label{fig:create_slab_submit}
\end{figure}

The starting point is the geometry of a molecule or a set of molecules uploaded in the AiiDA database.
A web-based interface allows the user to select a molecule and to build automatically a noble-metal substrate onto which the molecule can be positioned at will (see Fig.~\ref{fig:create_slab_submit}).

Once the geometry for a adsorbate-substrate system is available, either created with the mentioned interface or directly uploaded into the database, the equilibrium geometry can be obtained via submission of a geometry optimization workflow. 
The workflow automatically resubmits the calculation in case convergence has not been achieved within the queuing time limit on the compute infrastructure.

For specific substrate-adsorbate combinations, it is possible to select different levels of theory for the geometry optimization: \textit{Full DFT} will perform a DFT simulation for the whole substrate + adsorbate system; \textit{Mixed DFT (DFTB)} will treat the molecules at the DFT (DFTB \cite{elstner_self-consistent-charge_1998}) level and the substrate with EAM potentials \cite{foiles_embedded-atom-method_1986} and an empirical potential will couple the substrate to the molecule, mimicking van der Waals attraction and Pauli repulsion \cite{pignedoli_simple_2010}.
Once the equilibrium geometry of the system is obtained, a microscopy simulation can be performed, and the hybridization of the molecular levels with the substrate can be analyzed with the Scanning Probe Microscopy app described in Section~\ref{spm_app}.

\begin{figure}[H]
\centering\includegraphics[width=1.0\linewidth]{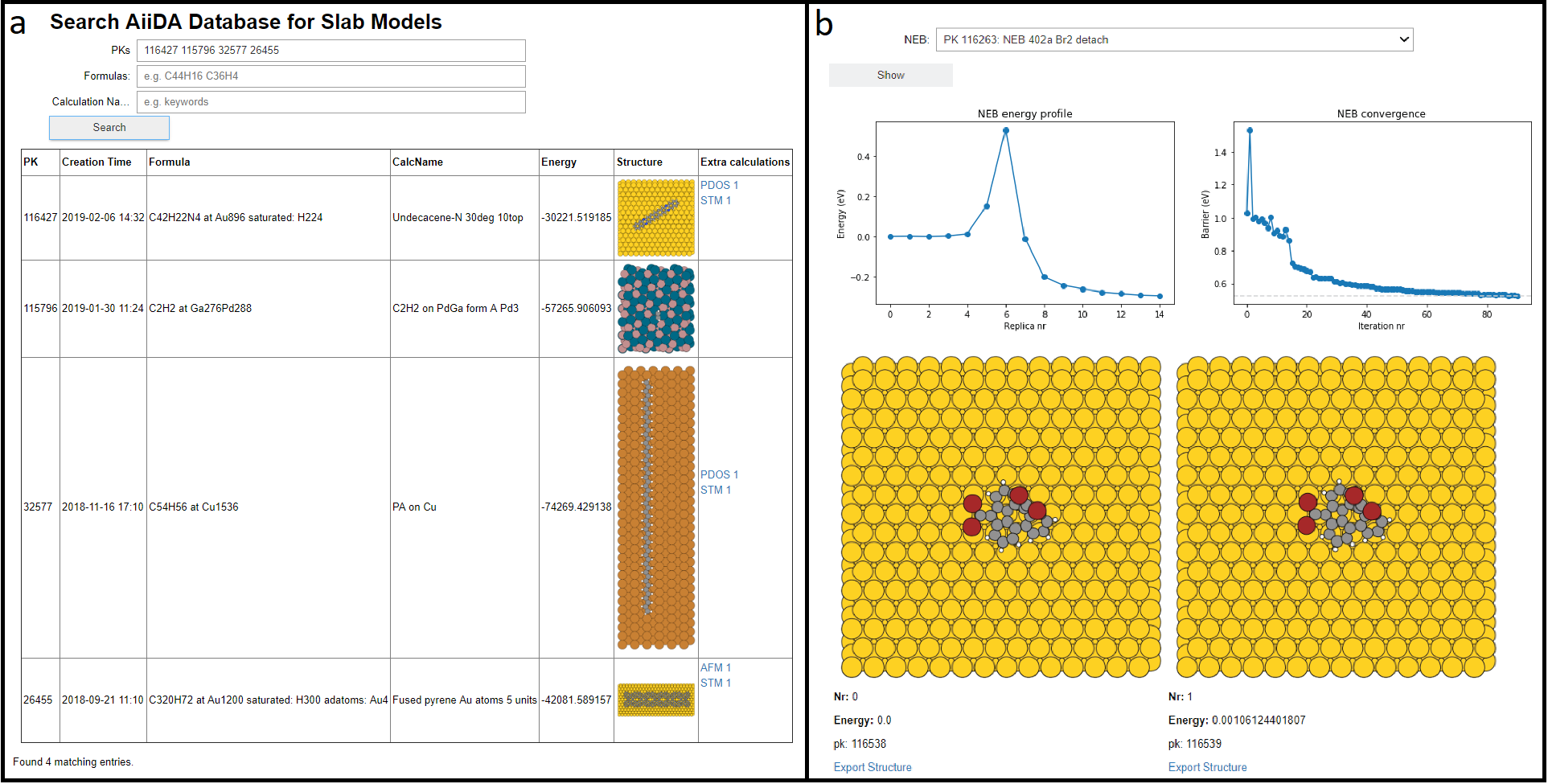}
\caption{Example user interfaces of the On-Surface Chemistry app. (a) Web-based interface to retrieve from the database the slab geometries that were optimized with the slab optimization workflow. Entries matching the search criteria are tabulated, together with selected slab properties and links to post-processing tools, such as PDOS and scanning probe microscopy. (b) Part of the web-based interface to visualize the results of a NEB calculation. In addition to the energy profile and convergence plots, a snapshot for each replica geometry is shown together with information on its energy and links to retrieve the geometry of the replica in xyz format.}
\label{fig:on_surf_app}
\end{figure}

The results of the slab optimization workflow can be accessed through a search interface.
In Fig.~\ref{fig:on_surf_app}a shows a list of optimized slabs retrieved from the database. For each geometry matching the search criteria, a thumbnail of the geometry together with the description of the calculation and the ``Total Energy'' of the structure are tabulated. An additional column displays links to additional calculations and tools available for a given geometry, if a PDOS analysis or a scanning probe calculation (STM or STS or AFM) has been performed with the Scanning Probe app (see section~\ref{spm_app}).

To investigate chemical reactions, such as dehalogenation of a GNR precursor or a polymerization process, we developed two workflows together with their interfaces: one to run a chain of constrained geometry optimizations and one to run Nudged-Elastic-Band (NEB) calculations.
Let us take as an example the case where we want to investigate a dehalogenation process, i.e. the removal of a bromine atom from the a GNR precursor molecule adsorbed on Au(111).
The molecule is shown in the NEB viewing interface, in Fig.~\ref{fig:on_surf_app}b.
To identify the transition state we use the NEB method once a suitable guess for the initial path of the reaction is known.
To create such a guess in complex scenarios, it is useful to perform a chain of constrained geometry optimizations, where a collective variable (CV) is defined within the atoms of the systems (e.g. the distance between to atoms, the angle between two bonds, or the angle between two planes identified by six atoms).
If we define as CV the C--Br distance, we can run a series of geometry optimization simulations where the value of the CV varies from the initial 1.9 {\AA} to a value (e.g. 4.0 {\AA}) corresponding to dehalogenation. The workflow expects as input a slab geometry stored in the AiiDA database, the definition of a CV (a subset of the CVs implemented in CP2K is defined) and the set of values that will have to be imposed to the CV. The app takes care of submitting a geometry optimization with the correct constraint value for each step, while using the converged geometry from the previous step as the initial geometry. Once the chain of constrained optimizations is completed, a visualization interface is able to retrieve the results, show a summary of all the geometries corresponding to the different constraint values, and allow to select specific geometries to set up the initial guess of a NEB calculation.

The NEB workflow then expects as input a slab geometry stored in AiiDA, a set of geometries sketching the initial guess for the reaction path and the total number of images to consider in the optimization of the reaction path. The app automatically takes care of running the NEB simulation and retrieving results. Visualization of the results is facilitated by means of a web interface that shows the energy profile along the reaction path, a convergence plot for the height of the barrier, and a list of chemical sketches for the geometries of the system along the path. A visualizer, shown in Fig.~\ref{fig:on_surf_app}b allows to follow interactively the geometry modifications along the reaction path.

\section{Detailed overview of the Scanning Probe Microscopy app}
\label{spm_app}

\begin{figure}[H]
\centering\includegraphics[width=.9\linewidth]{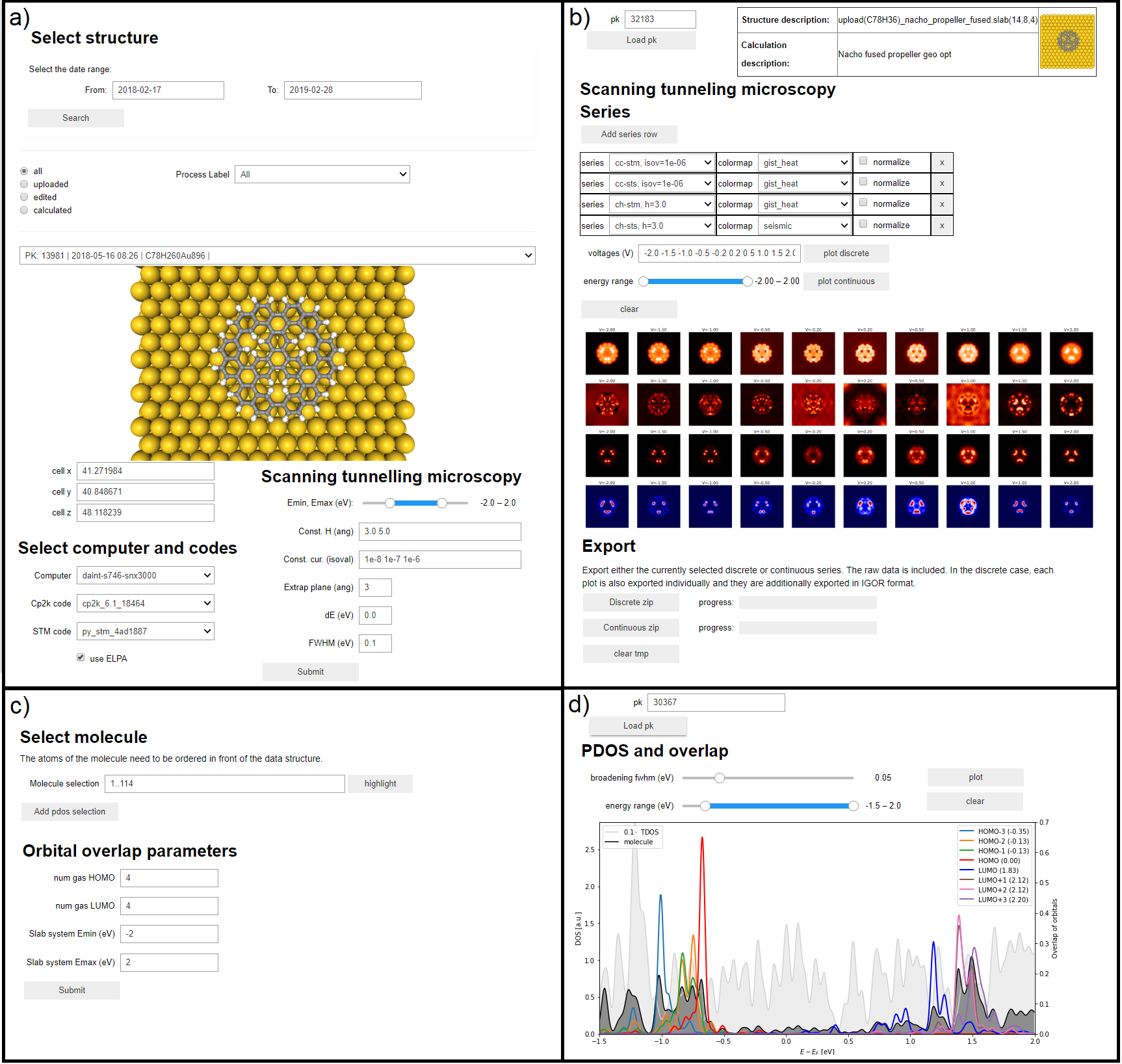}
\caption{Scanning Probe Microscopy app submission and viewing interfaces. (a) Submission interface of the STM calculation. It allows to specify as input any geometry in the AiiDA database; change its periodic cell, if needed; and select the remote computer and codes. Additionally the scanning tunneling microscopy parameters can be specified: bias voltage range as energy limits w.r.t.\@ the Fermi energy; constant-height plane position w.r.t.\@ the highest atom in \AA; constant current integrated local density of states isovalues; height at which the orbital extrapolation starts; energy differential for the spectroscopy series; and the full width at half maximum for the energy broadening of spectroscopy features. (b) Viewing interface of the STM calculation. It allows to visualize a series of images by specifying discrete bias voltage values or a continuous range of energies. In the example shown, a discrete series is visualized, which can be zoomed by double clicking. In the continuous case, a visualization box with a scroll bar appears. Additionally, any single image can be visualized and the selected visualizations can be exported as a ZIP file archive containing the images in PNG, IGOR Pro and raw data formats. (c) Elements of the submission interface of the orbital hybridization calculation. The structure input and code selection elements are equivalent to the STM case and are not shown in the panel. The interface allows to define the atoms of the molecule and additional selections where to project the DOS and the number of gas-phase orbitals to investigate the hybridization for. (d) Viewing interface for the orbital hybridization calculation. It allows for a flexible way to plot the PDOS and the orbital overlaps on the same figure.}
\label{fig:spm}
\end{figure}

The interface to submit a STM/STS workflow is shown on Fig.~\ref{fig:spm}a, where the user can specify the input geometry from the AiiDA database, the remote computer and codes, and the STM parameters.
The workflow performs a single point DFT calculation with CP2K to obtain the Kohn-Sham orbitals and the Hartree potential.
This output is then used as an input to a custom python toolkit~\cite{eimre_cp2k_2020}, which performs the STM/STS simulation within the Tershoff-Hamann approximation~\cite{Tersoff1985a,Tersoff1989} and extrapolates the electronic orbitals to the vacuum region in order to correct the wrong decay of the charge density due to the localized basis set~\cite{Tersoff1989}.
Constant-height and constant-current STM and STS are supported, which are respectively performed at specified planes of constant height (measured from the highest atom) or isovalues of integrated local density of states in the corresponding bias range.
The Kohn-Sham orbitals are broadened in energy with a Gaussian function with the specified full width at half maximum (FWHM) to match the experimental broadening in spectroscopy features.
The output interface, shown in Fig.~\ref{fig:spm}b, allows to visualize and compare all of the simulated STM/STS series corresponding to the chosen input parameters.
The images can be exported in PNG or IGOR Pro format or as raw data.

The Scanning Probe Microscopy app also includes a workflow to perform PDOS and orbital hybridization analysis on systems consisting of an adsorbate on a metal slab.
The interface to run this calculation is shown in Fig.~\ref{fig:spm}c and it allows the user to specify the input geometry, selections of atoms where to project the states, and the number of gas phase HOMO and LUMO orbitals for which to investigate the hybridization.
The workflow will run two single point DFT calculations with CP2K: for the whole system and for just the adsorbate (without changing the adsorption geometry).
The PDOS analysis is performed with CP2K for the calculation containing the whole system.
The orbital hybridization is investigated by performing the scalar product (i.e., the overlap) of the gas phase orbitals with the orbitals of the full system.
The result of this operation shows how much of the gas phase orbitals are preserved in the full system orbitals and their location in energy.
This overlap is performed with the custom python toolkit~\cite{eimre_cp2k_2020}, which takes the Kohn-Sham orbitals of both DFT calculations as input.
The viewing interface for this workflow is shown in Fig.~\ref{fig:spm}d, which displays the results for a circular porous nanographene.
The interface provides a flexible way to plot the PDOS and the orbital overlap results.
Each of the peaks in the PDOS on the adsorbate can be assigned to a single or to a combination of the gas phase orbitals.
Furthermore, this information can be used to assign the calculated spectroscopy mappings to the molecular orbitals of the adsorbate.

\section*{References}
\bibliographystyle{bib-style}
\bibliography{supplement}